\documentclass[acmsmall]{acmart}
\AtBeginDocument{%
  }

\setcopyright{cc}
\setcctype{by-nc-nd}
\acmDOI{10.1145/3797114}
\acmYear{2026}
\acmJournal{PACMSE}
\acmVolume{3}
\acmNumber{FSE}
\acmArticle{FSE086}
\acmMonth{7}
\acmSubmissionID{fse26maina-p3213-p}
\received{2025-09-12}
\received[accepted]{2025-12-22}

\usepackage{xspace}
\usepackage{multicol}
\usepackage{multirow}
\usepackage{graphicx}
\usepackage{subcaption}
\usepackage{array}
\usepackage{tcolorbox}
\usepackage{adjustbox}
\usepackage{enumitem}
\usepackage{booktabs, tabularx}
\usepackage[ruled, vlined, linesnumbered]{algorithm2e}

\newtcolorbox{rqbox}[1][]{
  colback=gray!5,
  colframe=gray!50,
  boxrule=0.5pt,
  arc=2pt,
  left=3pt, right=3pt,
  top=2pt, bottom=2pt,
  before upper={\textbf{#1: }},
}

\begin{document}

\renewcommand{\labelitemii}{$\circ$}
\newcommand{\fixme}[2][red]{\textcolor{#1}{FIXME: #2}}
\newcommand{\update}[2][blue]{\textcolor{#1}{#2}}
\newcommand{\addcite}[2][orange]{\textcolor{#1}{ADDCITE: #2}}
\newcommand{\name}{\textsc{Clotho}\xspace}

\newcommand{\fan}{failure@$N$\xspace}
\newcommand{\lohsv}{LOHS$_{var}$\xspace}
\newcommand{\gmmb}{GMM$_{b}$\xspace}
\newcommand{\sem}{SEM$_{ent}$\xspace}
\newcommand{\toke}{TOK$_{ent}$\xspace}
\newcommand{\tokp}{TOK$_{prob}$\xspace}

\newcommand{\odd}{\texttt{ODD-ADD}\xspace}
\newcommand{\gh}{\texttt{GH-TYPO}\xspace}
\newcommand{\json}{\texttt{JSON-FIX}\xspace}
\newcommand{\model}{\texttt{MODEL-EX}\xspace}
\newcommand{\pos}{\texttt{POS-TAG}\xspace}
\newcommand{\spell}{\texttt{SPELL-CHK}\xspace}
\newcommand{\syn}{\texttt{SYN-BUG}\xspace}
\newcommand{\topic}{\texttt{TOPIC-CLS}\xspace}

\title[\name: Measuring Task-Specific Pre-Generation Test Adequacy for LLM Inputs]{\name: Measuring Task-Specific Pre-Generation\\Test Adequacy for LLM Inputs}

\author{Juyeon Yoon}
\orcid{0000-0003-2706-1156}
\affiliation{%
  \institution{Korea Advanced Institute of Science and Technology}
  \city{Daejeon}
  \country{Republic of Korea}
}
\email{juyeon.yoon@kaist.ac.kr}

\author{Somin Kim}
\orcid{0009-0007-8917-7302}
\affiliation{%
  \institution{Korea Advanced Institute of Science and Technology}
  \city{Daejeon}
  \country{Republic of Korea}
}
\email{somin.kim@kaist.ac.kr}

\author{Robert Feldt}
\orcid{0000-0002-5179-4205}
\affiliation{%
  \institution{Chalmers University of Technology}
  \city{Gothenburg}
  \country{Sweden}
}
\email{robert.feldt@chalmers.se}

\author{Shin Yoo}
\orcid{0000-0002-0836-6993}
\affiliation{%
  \institution{Korea Advanced Institute of Science and Technology}
  \city{Daejeon}
  \country{Republic of Korea}
}
\email{shin.yoo@kaist.ac.kr}

\renewcommand{\shortauthors}{Juyeon Yoon, Somin Kim, Robert Feldt, Shin Yoo}

\begin{abstract}
Software increasingly relies on the emergent capabilities of Large Language Models (LLMs), from natural language understanding to program analysis and generation. Yet testing them on specific tasks remains difficult and costly: many prompts lack ground truths, forcing reliance on human judgments, while existing test adequacy measures typically rely on output uncertainty and thus are only available after full inference. A key challenge is to assess how useful a test input is in a way that reflects the demands of the task, ideally before even generating any output.
We introduce \name, a task-specific, pre-generation test adequacy measure that estimates input difficulty directly from LLM hidden states. Given a large pool of unlabelled inputs for a specific task, \name uses a Gaussian Mixture Model (GMM) to adaptively sample the most informative cases for human labelling. Based on this reference set the GMM can then rank unseen inputs by their likelihood of failure. In our empirical evaluation across eight benchmark tasks and three open-weight LLMs, \name can predict failures with a ROC-AUC of 0.716, after labelling reference sets that are on average only 5.4\% of inputs. It does so without generating any outputs, thereby significantly reducing LLM execution costs compared to output-based uncertainty or confidence measures. Comparison of \name and these post-generation adequacy measures shows that the two approaches complement each other.
Crucially, we show that adequacy scores learnt from open-weight LLMs transfer effectively to proprietary models, extending the applicability of the approach. When prioritising test inputs for proprietary models, \name increases the average number of failing inputs from 18.7 to 42.5 out of 100, compared to random prioritisation.
\end{abstract}


\maketitle

\section{Introduction}
\label{sec:intro}

The emergent capabilities of Large Language Models (LLMs) have enabled new types of software systems, from domain-specific natural language understanding~\cite{Kim2023ab} to autonomous coding and testing agents~\cite{Zhang2024aa,Yang2024ab,Yoon2024aa,Rondon2025wu}. As LLMs move beyond text generation into specific inference and reasoning tasks and see more widespread adoption, testing them becomes more critical and more costly. Yet the very qualities that make LLMs powerful also make them difficult to validate: their inner workings are opaque, and for many tasks we rely on them to solve, no ground truth is immediately available. 
Our goal is to improve task-specific LLM testing by \emph{prioritising which inputs to run and label first}, ideally exposing failures earlier and with less effort.

Prior work on test adequacy in Deep Neural Networks (DNNs) have focused on structural coverage~\cite{Pei2017qy,Ma2018aa} and distributional deviation from training data~\cite{Kim2019aa,Kim2021ct}. These test adequacy metrics enable us to prioritise manual labelling for new inputs~\cite{Kim2020zg}, or to guide automated input generation~\cite{Tian2018aa}. However, these approaches do not translate well to LLMs. First, LLM pre-training corpora are vast and heterogeneous~\cite{Gao2020aa,Laurencon2023ab}, making it infeasible to assess coverage or distributional density. Second, what matters in practice is not pre-training behaviour, it is rather how the model performs on a specific task as described by a fixed prompt\footnote{Technically a prompt is a prompt template describing a task, into which specific inputs are then inserted.}. 

Since adequacy cannot be judged against pre-training data, we can turn to signals derived from the model itself. Transformer-based LLMs~\cite{vaswani2017attention} encode input as embeddings, process it through stacked attention and feed-forward layers, and decode responses token by token. Within this structure, several post-hoc uncertainty measures have been proposed: semantic entropy to detect hallucinations~\cite{farquhar2024detecting}, self-consistency as a proxy for reliability~\cite{Wang2023aa}, and measures like token-level confidence and inference variability to flag input risk~\cite{huang2025look}. While these metrics are informative, they are all \emph{post-generation}. They exploit signals available only after output is produced, but thus require running full inference, often multiple times due to output non-determinism, which increases both computation time and cost. This limits their practicality for large-scale testing.

To tackle these challenges, we introduce \name\footnote{Clotho, the Greek goddess of fate, spins the thread of human life. The name suggests that our technique operates at the start of LLM inference, the pre-generation stage.}, a task-scoped adequacy measure designed around two key elements. First, it embeds testing in an \emph{active learning loop}~\cite{ren2021survey} that incrementally expands a small seed set of labelled inputs, reducing reliance on costly ground truth by focusing human effort where it matters most. Second, it operates \emph{pre-generation}, using only input embeddings of an LLM to estimate adequacy before any output is produced. This avoids the overhead of full inference while still identifying inputs likely to reveal failures.

Concretely, \name constructs a Gaussian Mixture Model (GMM) over the hidden states from the last token of passing inputs in labelled seed set. This surrogate model (i.e., predicting uncertainty in place of the target LLM) ranks unseen inputs by their likelihood-based ``surprise'', in the spirit of Surprise Adequacy~\cite{Kim2019aa,Kim2022hg}, with low-likelihood inputs treated as more challenging. The reference set is expanded iteratively: new labelling targets are chosen where the existing distribution is ambiguous, and also where they add diversity, so that human effort focuses on the most informative cases. To align with reference set expansion, the GMM adapts its number of components and dimensionality guided by cross-validation and estimates of active component usage.
Such iterative refinement enables the model to estimate input adequacy for unseen inputs more accurately. While pre-generation signals require internal hidden states, we show that scoring models trained on small open-weight LLMs transfer effectively to closed-weight proprietary models (e.g., GPT, Claude, Gemini), enabling cost-effective testing even with API-only access.

We validate \name across eight benchmark tasks and three open-weight LLMs (Llama, Mistral, Gemma). After running and labelling just 5.4\% of inputs, it achieves a ROC-AUC of 0.716 in predicting correctness on unseen inputs---on par with post-generation uncertainty metrics that require repeated inferences. When used to prioritise test inputs for proprietary models, \name surfaces 42.5 failures out of the top 100 inputs on average, compared to 18.7 of random selection, an increase of 126.8\%.

Our contributions are:

\begin{itemize}
    \item We propose \name, a task-specific, pre-generation adequacy measure that ranks inputs based on learnt density over hidden LLM states. \name adaptively constructs a reference set of labelled inputs, minimising the human labelling efforts. The reference set is used to model the distribution of passing inputs in the input-side activation space, which in turn allows us to predict whether an unseen input will produce a correct output.
    \item We conduct an empirical study across eight tasks and three open-weight LLMs, 
    showing strong prediction and prioritisation using limited labelling. 
    \item We demonstrate complementarity with post-generation uncertainty measures used for hallucination detection, such as internal hidden state variance and semantic entropy.
    \item We show how adequacy scores learnt on small models transfer to proprietary LLMs, enabling cost-effective testing even with API-only access.
\end{itemize}

Section~\ref{sec:exploratory} motivates and validates our pre-generation, task-scoped premise with an exploratory analysis of Transformer hidden states. Section~\ref{sec:approach} then describes \name in detail. Section~\ref{sec:exp_setup} lays out the details of our experimental setup, the result of which is presented in Section~\ref{sec:results}. Section~\ref{sec:discussion} discusses further research directions, and Section~\ref{sec:threats} lists threats to validity. Section~\ref{sec:related_work} presents related work and finally Section~\ref{sec:conclusion} concludes.

\section{Exploratory Study: Transformer Hidden States under Task-specific Prompting}
\label{sec:exploratory}

\name's core assumption is that both the type of task and the relative difficulty of each input for a task are reflected in the internal hidden states within the Transformer architecture, even \emph{before} any output token is produced. To verify the validity of these assumptions, several key questions need to be answered: (1) How do the Transformer internal states represent different tasks when the model is prompted with specific instructions? (2) Can we reliably distinguish potentially failing inputs from straightforward passing inputs based solely on Transformer internal states? (3) Which layers of the Transformer architecture provide the most useful hint for characterising input difficulty?

In this section, to explore the feasibility of leveraging hidden states to derive test adequacy metrics, we present preliminary studies that aim to answer these questions.

\begin{figure}
\begin{minipage}[t]{0.475\textwidth} %
    \begin{subfigure}[t]{0.475\textwidth}
        \includegraphics[draft=false, width=\textwidth]{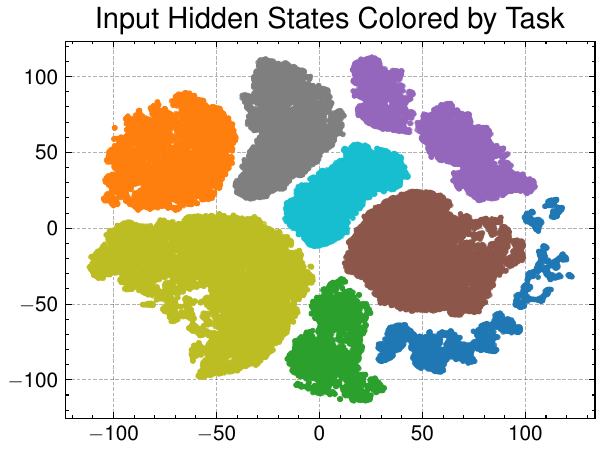} 
        \caption{Task Clusters}\label{fig:task-clusters}
    \end{subfigure}
    \hfill
    \begin{subfigure}[t]{0.475\textwidth}
        \includegraphics[draft=false, width=\textwidth]{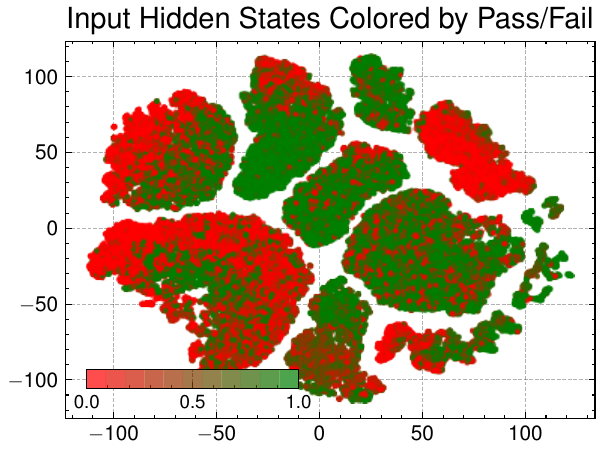} 
        \caption{Pass/Fail Separation}\label{fig:pass-fail-separation}
    \end{subfigure}
    \caption{Visualisation on Llama 8B internal state space using inputs for all studied tasks}
    \label{fig:exploratory-state_visualization}
\end{minipage}
\hfill
\begin{minipage}[t]{0.475\textwidth} %
    \begin{subfigure}[t]{0.475\textwidth}
        \includegraphics[draft=false, width=\textwidth]{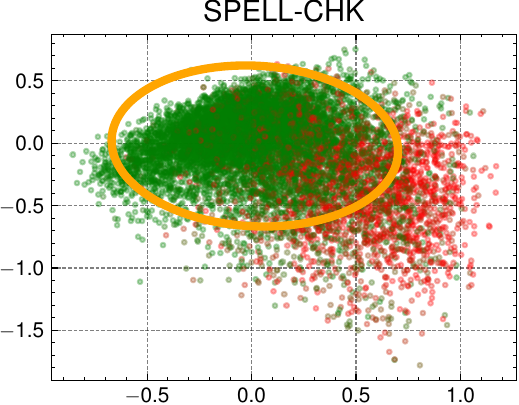} 
        \caption{Within-Task}\label{fig:within-task-pca}
    \end{subfigure}
    \hfill
    \begin{subfigure}[t]{0.475\textwidth}
        \includegraphics[draft=false, width=\textwidth]{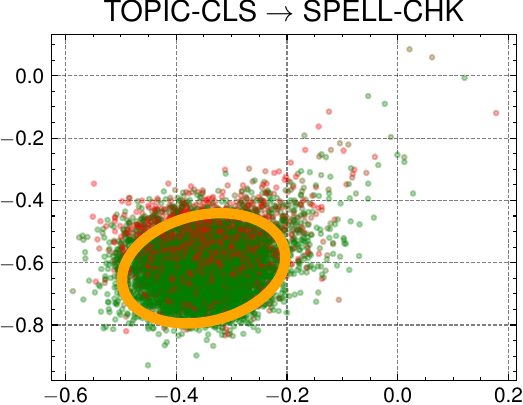} 
        \caption{Cross-Task}\label{fig:cross-task-pca}
    \end{subfigure}
    \caption{Pass/fail separation of \spell inputs in projected PCA space of passing inputs}
    \label{fig:cross-task-visualization}
\end{minipage}
\end{figure}

\subsection{Modelling Representative Cases in Task-specific Internal State Subspace}
\label{sec:exploratory_task_specific_internal_state}

Figures~\ref{fig:task-clusters} and~\ref{fig:pass-fail-separation} plot activation vectors from a hidden layer of Llama 3.1 8B for inputs from eight natural and programming language tasks studied in this paper (see Section~\ref{sec:exp_setup_task_dataset}). We reduced the vectors to two dimensions using t-SNE~\cite{maaten2008visualizing}. In Figure~\ref{fig:task-clusters}, colours mark tasks; in Figure~\ref{fig:pass-fail-separation}, colours represent the proportion of correct outputs across ten repeated generations, with greener shades indicating more correct outputs and redder shades indicating more incorrect outputs. It shows that the eight tasks form well-separated clusters in the internal representation space of inputs. To ensure that this structure is not an artefact of t-SNE exaggeration, we compute the silhouette score in the 10-dimensional PCA space, obtaining a value of 0.9579, which indicates exceptionally strong clustering. Consistently, intra-cluster cosine distances are very small (mean = 0.0265, max = 0.0490), whereas inter-cluster distances are substantially larger (mean = 1.1272).\footnote{The smallest inter-cluster distance (0.2617) occurs between GH-TYPO and SPELL-CHK, which is intuitively reasonable given their shared focus on misspellings and typos.}

More importantly, Figure~\ref{fig:pass-fail-separation} shows a noticeable separation between green and red regions even in the highly reduced 2D space, suggesting that internal input representations may contain information that distinguishes easier inputs from the more challenging ones. We further examined how input distributions vary depending on their pass/fail outcomes: we fitted Principal Component Analysis (PCA)~\cite{abdi2010principal} on the activations of inputs for which the majority of repeated generations are correct (hereafter referred to as \emph{passing} inputs) from the \spell task,
and then transformed all inputs from the same \spell task, including failing ones, into the fitted space (Figure~\ref{fig:within-task-pca}). We observe a clear pattern where failing inputs deviate from the dense region of passing inputs. However, we note that this distributional deviation is task-specific. 

Figure~\ref{fig:cross-task-pca} presents the same \spell inputs, but mapped into the PCA space fitted with the passing inputs of a different task, \topic. In this cross-task projection, the separation between passing and failing inputs become notably weaker compared to the within-task projection. Examining this effect through the Fisher Discriminant Ratio (FDR)~\cite{fisher1936use} between passing and failing groups, we find that within-task PCA consistently yields stronger discriminability than cross-task PCA, including SPELL-CHK (0.736 vs. 0.314), ODD-ADD (0.593 vs. 0.246), and JSON-FIX (0.101 vs. 0.015). From this observation, we posit that distributional deviation in hidden state space of LLMs can indeed provide a useful signal for assessing test adequacy, but only when analysed in a \emph{task-specific} way. PCA was used here because t-SNE does not transfer across different data samples.

Finally, Figure~\ref{fig:task-clusters} also shows that some tasks yield multi-modal distributions, with several distinct clusters instead of a single dense region. This means that out-of-distribution detection cannot be simply reduced to measuring distance from the centre of one unimodal distribution. It calls instead for models that can capture multi-modality~\cite{Kim2021ct}.

\begin{figure}[t]
    \centering
    \includegraphics[width=\linewidth]{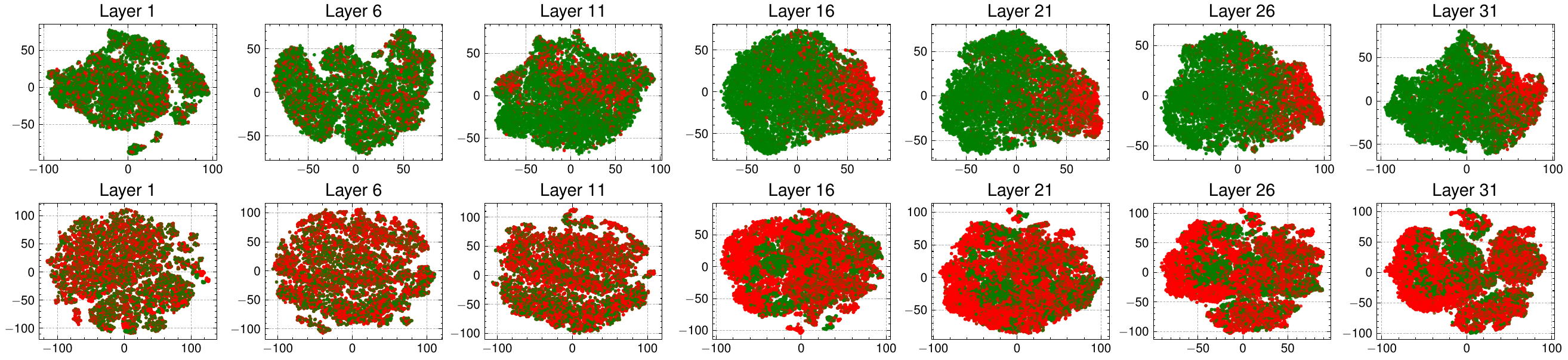}
    \caption{Input hidden states of Llama 8B from two prompted tasks (one per line) for varying depths of layers.}
    \label{fig:exploratory/layer_selection}
\end{figure}

\subsection{Extracting Internal Hidden Vectors from Transformer Architecture}
\label{sec:exploratory_token_layer_choice}
We now turn to the problem of extracting internal states from LLMs. Most contemporary LLMs use the Transformer architecture, which maps each token to an embedding vector and then processes it through a stack of Transformer blocks. Each block contains a self-attention sublayer and a feed-forward sublayer, each wrapped with residual connections. At every layer, the model maintains a residual stream: a continuous vector representation that accumulates contextual information. These vectors are what we refer to as the \textit{hidden states}. Defining a hidden space for measuring test adequacy therefore requires two choices: which input token to extract from, and which layer's residual vector to use.

The token choice is straightforward. In decoder-only models, the hidden state at the final token position encodes the entire prefix, thanks to the autoregressive design. We consider two candidate positions: the last token of the input prompt and the last token of the generated output. The final input token is preferable, as we saw earlier that its hidden states can yield clear separations by task difficulty (Figure~\ref{fig:pass-fail-separation}), while it also would avoid the computational overhead of generating outputs. For this reason, we focus on the \emph{final input token}.

Layer choice is less obvious. Prior work shows that middle-to-late layers capture semantic abstractions more effectively, for example, in hallucination detection~\cite{kossen2024semantic, azaria-mitchell-2023-internal} and adversarial input identification~\cite{zhou2024investigating}. We verify this in our setting by examining hidden states of the final input token across all layers in Llama3.1 8B (32 layers in total), using one natural language task (\spell) and one programming task (\syn). Figure~\ref{fig:exploratory/layer_selection} illustrates the results for a subset of the layers, visualised with t-SNE, after first reducing the dimensionality to 10 using PCA for computational efficiency. Early layers show little distinction between passing and failing cases. In middle and later layers, however, separation emerges, with inputs clustering more distinctly by outcome. The effect seems strongest in the \emph{latter half of the model's depth}.

Guided by this evidence, we extract hidden states from the layer at two-thirds of the model's total depth, corresponding to the last input token. We denote these as LIHS (\underline{L}ast-token \underline{I}nput \underline{H}idden \underline{S}tates). For comparison, in our empirical work, hidden states of the last output token are denoted LOHS (\underline{L}ast-token \underline{O}utput \underline{H}idden \underline{S}tates).

\section{Approach}
\label{sec:approach}
\name formulates the input-level test adequacy as the degree of distributional deviation~\cite{Kim2019aa,Kim2022hg}. To align the \emph{reference} distribution with the given task, without relying on LLM training data, \name adaptively models the distribution of inputs the target LLM can successfully handle. This is achieved by fitting a Gaussian Mixture Model (GMM) to hidden state embeddings (LIHS) of passing inputs, expanded through strategic sampling. This section describes details of the approach.

\subsection{Measuring Test Adequacy as Distributional Deviation}
\label{sec:approach_sampling}

Given a labelled reference set of inputs, \name estimates probability density of \emph{passing} inputs in the space of LLM hidden states. More concretely, we extract LIHS of passing inputs and apply GMM to capture their distribution. The use of GMM is motivated by the fact that the distribution of LIHS may not be unimodal (see Section~\ref{sec:exploratory_task_specific_internal_state}).

\begin{algorithm}[t]
\caption{\name}
\label{alg:clotho}
\KwIn{a set of unlabelled inputs, $\mathcal{U}$, the initial reference set of labelled inputs, $\mathcal{R}_0$, target reference set size, $N$, batch size, $B$, exploration ratio, $\alpha$}
\KwOut{an input distribution model $\mathcal{M}_T$ trained on $\mathcal{R}_\text{pass}$ 
       with $|\mathcal{R}| = N$}

$\mathcal{R} \leftarrow \mathcal{R}_0$ \;
$t \leftarrow 0$ ; \tcp{iteration count}
\While{$|\mathcal{R}| < N$}{
    Adapt $\mathcal{M}_t$ with best representation dim. and \# of mixture components w.r.t. $\mathcal{R}$ \;
    $\mathcal{R}_\text{pass} \leftarrow \{x \in \mathcal{R} \mid \text{label}(x)=\text{pass}\}$ \;
    Fit GMM $\mathcal{M}_t$ on $\mathcal{R}_\text{pass}$ \;
    
    \For{$x \in \mathcal{U}$}{
        $s_\text{exploit}(x) \leftarrow$ entropy of component responsibilities under $\mathcal{M}_t$ \;
    }

    $B_\text{explore} \leftarrow \lfloor \alpha B \rfloor $, $B_\text{exploit} \leftarrow B - B_\text{explore}$ \;
    
    $C_\text{exploit} \leftarrow$ the $B_\text{exploit}$ inputs with the highest $s_\text{exploit}$ values \;
    $C_\text{explore} \leftarrow \emptyset$ \;
    \For{$j = 1$ to $B_\text{explore}$}{
        For each $x \in \mathcal{U} \setminus (C_\text{exploit} \cup C_\text{explore})$, \\
        compute $d(x) = \min_{y \in \mathcal{R} \cup C_\text{exploit} \cup C_\text{explore}} \text{Dist}(x,y)$ \;
        Pick $x^* = \arg\max_x d(x)$ \;
        $C_\text{explore} \leftarrow C_\text{explore} \cup \{x^*\}$ \;
    }
    $C \leftarrow C_\text{exploit} \cup C_\text{explore}$ \;
    
    Execute and obtain human labels for inputs in $C$ \;
    $\mathcal{R}\leftarrow \mathcal{R} \cup C$, $\mathcal{U} \leftarrow \mathcal{U} \setminus C$, $t \leftarrow t + 1$ \;
    
}
\Return $\mathcal{M}_t$
\end{algorithm}

Algorithm~\ref{alg:clotho} outlines the procedure for iteratively sampling and constructing the reference set, and fitting a GMM on the passing reference inputs until the target reference set size (labelling budget) is reached. The GMM is always trained on the subset of reference inputs labelled as ``pass'' ($\mathcal{R}_\text{pass}$, line 5-6). We define a passing input as one that produces correct output in the majority of repeated runs. While alternative definitions are possible, this ensures that the learnt distribution captures inputs that the LLM under test can reliably handle, mitigating randomness.

In the main loop (lines 7-19), \name expands the reference set $\mathcal{R}$. We cast this as an active learning process~\cite{Cohn1996wo}, where inputs are iteratively selected for labelling. Each round targets inputs that maximise information gain using two complementary strategies: exploitation, which sharpens the boundaries of the known hidden state space, and exploration, which extends coverage into under-represented regions. To balance these goals, \name divides the batch into $B_{exploit}$ and $B_{explore}$ (line 9).
Exploitation is guided by the GMM's component assignment entropy on unlabelled inputs (lines 8 and 10). High entropy signals uncertainty about which component an input belongs to, so labelling the most uncertain inputs helps refine the model's understanding of the distribution. Exploration, in contrast, is driven by diversity: \name selects inputs with the greatest minimum Euclidean distance from those already in $\mathcal{R}$ (line 14), following the greedy max-min strategy of Adaptive Random Testing (ART)~\cite{chen_adaptive_2005}. Larger distances ensure broader coverage of the hidden state space. Since no prior knowledge of the latent space is available, \name weights exploitation and exploration equally, setting $\alpha = 0.5$: we may consider adaptive fine-tuning in the future.

Using Algorithm~\ref{alg:clotho}, \name produces a Gaussian Mixture Model, $\mathcal{M}_T$, that estimates the probability density of hidden states from passing inputs for a given task:

\[
\mathcal{L}_{\text{pass}}(x) = p_\theta\!\left(h(x)\right), 
\quad h(x) \text{: hidden state representation (LIHS)}
\]

The density $p_\theta\!\left(h(x)\right)$ captures how plausible an input is relative to the reference distribution of LIHS from passing cases: higher values indicate that $x$ falls within regions commonly associated with correct outputs. From this density, we can compute Likelihood-based Surprise Adequacy (LSA)~\cite{Kim2019aa}, defined as the negative log likelihood:

\[
\text{LSA}(x) = - \log \, p_\theta\!\left(h(x)\right)
\]

A higher LSA means the input is more surprising compared to the reference distribution and therefore more likely to trigger a failure. LSA can thus be used either to prioritise human labelling efforts or to identify more challenging inputs for testing.

\subsection{Adaptive Modelling of Passing Input Distribution via GMM}
\label{sec: adaptive_modelling}

In Algorithm~\ref{alg:clotho}, the model $\mathcal{M}_t$ is iteratively updated as the reference set expands. To improve model accuracy, we adapt two key parameters during this process: the dimensionality of the LIHS representations used by the GMM and the number of mixture components in the model. Both adaptations are performed at line 4 of Algorithm~\ref{alg:clotho}.

\subsubsection{Dimensionality Selection}

Training a GMM with limited data benefits from compact input representations; otherwise, training can become slow and inefficient. Our analysis shows that raw residual hidden state vectors are often too high-dimensional for the GMM to handle effectively, and must be reduced. Moreover, we have noted that the optimal dimensionality varies across tasks. To address this, we apply PCA to project raw hidden states into a lower-dimensional latent space and adaptively determine the ideal number of dimensions. Starting from 10 dimensions, \name explores two alternatives at each update step: one with 10 more and one with 10 fewer dimensions (lower bound is set to five to avoid collapse). It then performs 3-fold cross-validation on the reference set, based on the rank correlation between the GMM predicted likelihood of passing and the actual pass rate observed from a set of multiple runs of each input. The dimensionality with the highest mean correlation across folds is chosen for the next iteration, since a stronger correlation indicates a more faithful latent representation.

\subsubsection{Component Adaptation}

A GMM requires the number of mixture components, the Gaussian distributions combined to represent the data, as a hyperparameter. Ideally, this matches the number of task-specific hidden state clusters (see Section~\ref{sec:exploratory_task_specific_internal_state}), but that information is usually unavailable in advance. To address this, we adapt the number of components using a perplexity-based heuristic that measures how evenly the GMM distributes probability mass across its components. Given mixture weights $\mathbf{w} = (w_1, \dots, w_K)$, for a $K$ component GMM, we define component perplexity as

\[
\text{Perplexity}(\mathbf{w}) 
= \exp\!\left( - \sum_{k=1}^K w_k \log w_k \right),
\quad \text{where } \mathbf{w} = (w_1, \dots, w_K).
\]

\name starts with five mixture components. If perplexity shows that only a few components dominate, we reduce the count by one. If it indicates that many components are actively contributing, we increase the count by one, up to a maximum of 50.

\section{Experimental Setup}
\label{sec:exp_setup}
We present our experimental setup in this section.

\subsection{Research Questions}

Our empirical evaluations are designed to answer the following research questions.

\begin{itemize}
\item \textbf{RQ1.} How accurately does \name model the latent space of passing inputs as the reference set expands?
If the modelling is accurate, the predicted likelihood of \name would strongly correlate with the actual passing rates of unseen inputs. We answer RQ1 primarily by comparing the Spearman rank correlation computed for \name and other pre-generation baseline approaches that can estimate the likelihood of each input producing the correct output. Further, we also investigate how much the input sampling method of \name affects the correlation.

\item \textbf{RQ2.} How effective is \name at prioritising nontrivial test cases showing high failure rates for efficient testing? To answer RQ2, we predict pass/fail for inputs based on the predicted likelihood of \name and report ROC-AUC. We also report \fan, i.e., the number of failing inputs out of top $n$ inputs when sorted according to model score: higher \fan would indicate the used score can effectively prioritise failure inducing inputs.

\item \textbf{RQ3.} How do \name's adequacy scores complement existing post-generation uncertainty and confidence metrics? We answer RQ3 by comparing Spearman rank correlation of \name to internal state variance, semantic entropy, token probability, and token entropy. 

\item \textbf{RQ4.} Do adequacy scores from \name transfer from open-weight models to closed-weight proprietary models as valid indicators of input difficulty? We address RQ4 by prioritising inputs with \name trained on smaller open-weight LLMs and then evaluating that prioritisation on larger closed-weight models, including GPT, Claude, and Gemini.
\end{itemize}

\subsection{Target Tasks and Dataset}
\label{sec:exp_setup_task_dataset}

Table~\ref{tab:exp_setup_task_dataset} lists the eight tasks, their prompts and input datasets used in our evaluation. We curate these datasets according to three criteria:

\begin{itemize}
\item\textbf{Data Volume:} We focus on datasets with sufficiently large number of inputs, as validating the generalisability of the input distribution modelled by \name to unseen test inputs requires ample data. This prevents the use of smaller benchmarks consisting of only hundreds of samples, which are generally designed for broad LLM evaluation rather than for constructing task-specific test suites.
\item\textbf{Clear Ground Truths:} Many general Natural Language Understanding (NLU) benchmarks, such as question answering and summarisation, yield open-ended outputs that are hard to validate both automatically and objectively. We instead focus on tasks with unambiguous ground truth labels.
\item\textbf{Separation of Template and Inputs:} Since \name targets specific tasks rather than open-ended dialogue, we use datasets with a clear separation between the task template (instructions) and the input (the variable part, per LLM request). This mirrors realistic use cases already supported by popular development frameworks for LLM applications~\cite{langchain_prompt_template,sharma2025promptpex}.
\end{itemize}

\begin{table}[ht]
\centering
\scriptsize
\caption{Summary of Tasks, Sources, and Verification Methods Used in Our Experiments}
\label{tab:exp_setup_task_dataset}
\resizebox{\linewidth}{!}{%
\begin{tabular}{lrlll}
\toprule
\textbf{Task ID} & \textbf{\# Inputs} & \textbf{Template} & \textbf{Input} & \textbf{Aim} \\

\midrule
\texttt{ODD-ADD} & 6,000 & \href{https://www.promptingguide.ai/prompts/mathematics/odd-numbers}{P.E. Guide}~\cite{PromptEngineering.ai:sy} & Randomly Sampled Numbers & To find sum of only odd numbers \\

\midrule
\texttt{GH-TYPO} & 10,000 & Ours & \href{https://github.com/mhagiwara/github-typo-corpus}{GitHub Typo Corpus}~\cite{hagiwara2019github} & To fix typos extracted from GitHub commits \\

\midrule
\texttt{JSON-FIX} & 6,563 & Ours & Synthetic JSON + Bugs~\cite{github:jsonrepair} & To repair invalid JSON \\

\midrule
\texttt{MODEL-EX} & 9,810 & \href{https://www.promptingguide.ai/prompts/information-extraction/extract-models}{P.E. Guide}~\cite{PromptEngineering.ai:sy} & \href{https://huggingface.co/datasets/CShorten/Last-Week-on-ML-ArXiv}{ML-ArXiv}~\cite{huggingface:arxiv_abstracts_dataset}, Synthetic Abstracts & To extract ML model names from paper abstracts \\

\midrule
\texttt{POS-TAG} & 15,359 & \href{https://github.com/microsoft/promptpex}{PromptPex~\cite{sharma2025promptpex, schnabel2024symbolic}} & \href{https://github.com/UniversalDependencies/UD_English-EWT}{UD\_English-EWT}~\cite{silveira14gold} & To detect Part-of-Speech tags \\

\midrule
\spell & 10,000 & Ours & \href{https://wordnet.princeton.edu/}{WordNet sentences}~\cite{fellbaum1998wordnet} + \href{https://titan.dcs.bbk.ac.uk/~roger/corpora.html}{Misspellings}~\cite{mitton1980birkbeck} & To fix misspelt words \\

\midrule
\texttt{SYN-BUG} & 20,518 & \href{https://arxiv.org/abs/2406.15325}{BICS-Dataset}~\cite{lee2024bug} & \href{https://github.com/HammingHQ/bug-in-the-code-stack}{Python Syntactic Bug}~\cite{lee2024bug} & To find code lines with syntactic bugs \\

\midrule
\texttt{TOPIC-CLS} & 7,600 & \href{https://github.com/microsoft/promptpex}{PromptPex~\cite{sharma2025promptpex}} & \href{https://huggingface.co/datasets/fancyzhx/ag_news}{AG News}~\cite{Zhang2015CharacterlevelCN} & To classify topics of news articles \\

\bottomrule
\end{tabular}
} %
\end{table}

The prompt templates for \odd, \model are taken from Prompt Engineering Guide~\cite{PromptEngineering.ai:sy}, \pos and \topic from PromptPex~\cite{sharma2025promptpex}, and \syn from the BICS benchmark. The templates for \gh, \json, and \spell are written by authors to fit the corresponding tasks and datasets. For \model, we use paper abstracts from ML-ArXiv dataset~\cite{huggingface:arxiv_abstracts_dataset} but also synthesise abstracts that use real ML model names using GPT-4o. All prompt templates, input data, as well as scripts used to synthesise inputs, are available from our replication package.

\begin{table}[ht]
    \centering
    \caption{Pass rates ($p=0.5$) of the studied open-weight LLMs on the eight tasks}
    \label{tab:dataset_failure_rate}
    \resizebox{0.9\textwidth}{!}{%
    \begin{tabular}{lrrrrrrrr}
    \toprule
    & \odd & \gh & \json & \model & \pos & \spell & \syn & \topic \\
    \midrule
    Gemma 2 9B & 0.59 & 0.63 & 0.79 & 0.66 & 0.79 & 0.88 & 0.40 & 0.81 \\
    Llama 3.1 8B & 0.57 & 0.48 & 0.67 & 0.52 & 0.74 & 0.80 & 0.34 & 0.77 \\
    Mistral 7B & 0.08 & 0.29 & 0.31 & 0.49 & 0.68 & 0.77 & 0.28 & 0.75 \\
    \bottomrule
    \end{tabular}
    } %
\end{table}

\subsection{Models}
\label{sec:models}

To evaluate \name, we use three widely adopted Open-weight LLMs (OLMs): Llama3.1 8B~\cite{llama3modelcard}, Gemma 2 9B~\cite{gemma_2024}, and Mistral 7B~\cite{Jiang2023fv}. For hidden state extraction from LIHS and LOHS, we select layers at roughly two-thirds of model depth, guided by our preliminary analysis in Section~\ref{sec:exploratory_token_layer_choice} and prior work~\cite{kossen2024semantic,azaria-mitchell-2023-internal}. Table~\ref{tab:dataset_failure_rate} reports the pass rates of these models on our tasks: an input counts as passing if it yields the correct output in a majority of generations, more than five out of ten runs.
For RQ4, we additionally include three representative proprietary LLMs: GPT-4o mini\footnote{\url{https://openai.com/index/gpt-4o-mini-advancing-cost-efficient-intelligence/}}, Claude Haiku\footnote{\url{https://www.anthropic.com/claude/haiku}}, and Gemini 2.5 Flash Lite\footnote{\url{https://deepmind.google/models/gemini/flash/}}. These serve as targets for testing whether adequacy scores learnt from OLMs transfer effectively to closed-weight ones.

\subsection{Baselines}
\label{sec:baselines}

We compare \name against two categories of baseline metrics: pre-generation and post-generation.

\subsubsection{Pre-generation Adequacy Metrics}

We first consider baselines that, like \name, estimate test adequacy before any output is generated: \emph{Sequence Log Likelihood} (SLL), \emph{Mahalanobis Distance-based Surprise Adequacy} (MDSA)~\cite{Kim2020zg}, and \emph{Base Gaussian Mixture Model} (\gmmb). SLL is the average log-likelihood of input tokens computed using the LLM under test. It requires no labelled data and serves as a simple, label-free adequacy baseline. MDSA computes the Mahalanobis Distance~\cite{De-Maesschalck2000cj} between a new input and the distribution of inputs in the randomly sampled reference set. By assuming a unimodal distribution, it cannot capture the more complex structures observed in input space (see Section~\ref{sec:exploratory}). Finally, \gmmb is an ablated version of \name that omits both the balanced sampling strategy (Section~\ref{sec:approach_sampling}) and adaptive modelling (Section~\ref{sec: adaptive_modelling}), instead fixing the number of components at five and the latent dimensionality at 10, which is the initial parameters of \name.

\subsubsection{Post-generation Adequacy Metrics}

Since \name essentially predicts the correctness of the output induced by the given input, we can compare \name to post-generation uncertainty metrics: \emph{Token Probability} (\tokp) \& \emph{Token-Entropy} (\toke)~\cite{manakul2023selfcheckgpt}, \emph{Semantic Entropy} (\sem)~\cite{farquhar2024detecting}, and \emph{Last Output-token Hidden State Variance} (\lohsv). \tokp and \toke are widely used confidence measures computed from a generated sequence: they can be computed by taking the token level likelihood and entropy at each decoding step of a single output. Following prior work, we derive sentence-level scores by averaging token log-probabilities and entropies over the sequence. \sem requires multiple outputs from the same prompt, and computes the entropy across semantic clusters formed by the embeddings of outputs. We cluster semantically equivalent outputs among 10 generations using a DeBERTa-based NLI model~\cite{he2020deberta}, before computing entropy over the resulting clusters. Finally, \lohsv captures the variance among Last Output-token Hidden States by taking the trace of the covariance matrix of the collected LOHS vectors. This score reflects how much the model's internal state diverge when producing different responses to the same prompt, thereby serving as an internal measure of generation variance.

\subsection{Evaluation Protocols and Metrics}

Here, we describe the process of experimentation as well as the evaluation metrics used.

\subsubsection{Reference Sets and Implementation of \name}

\name requires a small initial reference set as a starting point. For each task, we provide 10 inputs that mostly yield correct outputs as the initial reference set: these are manually crafted and not from the datasets listed in Table~\ref{tab:exp_setup_task_dataset}. This is intended to simulate a realistic use case, in which a developer has a small set of \emph{working} inputs for a prompt, before starting a more rigorous testing process. One exception is \syn, for which we find it difficult to craft inputs that reliably pass with our studied OLMs (i.e., inputs with passing rate greater than 0.5), however simple the input is. Consequently, the initial reference set for \syn contains fewer than ten reliably passing inputs. The difficulty of this task is also noted in the original paper~\cite{lee2024bug}. Note that the same initial reference set is used by \name, MDSA, and \gmmb for RQ1.

\name is implemented using \texttt{GaussianMixture} from \texttt{sklearn.mixture} with full covariance matrices~\cite{Pedregosa2011tk}. Training terminates when the log-likelihood improvement falls below $10^{-3}$ or after 500 iterations, whichever comes first. Each iteration uses a batch size of 10, and the maximum target reference set size is 500. To account for randomness in GMM training, the entire process is repeated with ten different random seeds.

\subsubsection{Evaluation Metrics}

We measure the Spearman rank correlation between the predicted likelihoods and the actual pass rates observed from multiple runs of the same input (we repeat ten times). Spearman rank correlation measures how strongly two sets of ranks are correlated with each other. For RQ2, we report ROC-AUC for binary failure prediction, labelling inputs with pass rate greater than 0.5 as passing. We apply the Mann-Whitney U test to assess statistical significance, and measure \fan, the number of failing inputs among the top $N$ ranked by the score.

\section{Results}
\label{sec:results}

In this section, we present our evaluation results.

\subsection{Failure Prediction Performance (RQ1)}

We begin by evaluating how well \name can model the distribution of passing inputs in the latent space, and compare \name to other approaches that are applicable at the pre-generation stage.

\subsubsection{RQ1-1: Score Correlation with Actual Pass Rate Observed} 

\begin{table}[ht]
\caption{Comparison of pre-generation metrics across reference set sizes (Spearman rank correlation between predicted likelihoods and observed pass rates, higher values indicates better prediction performance).}
\label{tab:compare_metrics}
\resizebox{0.9\textwidth}{!}{%
\begin{tabular}{lc|rrrr|rrrr|rrrr}
\toprule
\multicolumn{2}{c}{}  & \multicolumn{4}{c}{Gemma 2 9B}
 & \multicolumn{4}{c}{Llama 3.1 8B}
 & \multicolumn{4}{c}{Mistral 7B} \\
\midrule
Task ID & N & \name & MDSA & \gmmb & SLL & \name & MDSA & \gmmb & SLL & \name & MDSA & \gmmb & SLL \\\midrule
\multirow[c]{3}{*}{\odd}  & 100& \textbf{0.451}& 0.190         & 0.082         & \multirow{3}{*}{-0.201}& \textbf{0.558}& 0.333         & 0.087         & \multirow{3}{*}{-0.336}& \textbf{0.371}& 0.261 & 0.215         & \multirow{3}{*}{-0.434} \\
                          & 300& \textbf{0.407}& 0.212         & 0.157         &                        & \textbf{0.542}& 0.312         & 0.207         & & \textbf{0.389}& 0.317 & 0.081         &  \\
                          & 500& \textbf{0.434}& 0.184         & 0.158         &                        & \textbf{0.551}& 0.239         & 0.173         & & \textbf{0.432}& 0.356 & 0.089         &  \\\midrule
\multirow[c]{3}{*}{\gh}   & 100& 0.302         & 0.451         & \textbf{0.472}& \multirow{3}{*}{-0.046}& \textbf{0.431}& 0.430         & 0.392         & \multirow{3}{*}{-0.039}& \textbf{0.195}& 0.125 & 0.126         & \multirow{3}{*}{-0.037} \\
                          & 300& 0.305         & \textbf{0.468}& 0.364         &                        & 0.392         & \textbf{0.469}& 0.439         & & \textbf{0.189}& 0.143 & 0.180         &  \\
                          & 500& 0.440         & \textbf{0.455}& 0.330         &                        & 0.456         & \textbf{0.482}& 0.434         & & \textbf{0.188}& 0.129 & 0.138         &  \\\midrule
\multirow[c]{3}{*}{\json} & 100& \textbf{0.294}& 0.243         & 0.200         & \multirow{3}{*}{-0.005}& \textbf{0.289}& 0.153         & 0.072         & \multirow{3}{*}{0.061} & \textbf{0.443}& 0.354 & 0.253         & \multirow{3}{*}{0.032} \\
                          & 300& \textbf{0.359}& 0.238         & 0.218         &                        & \textbf{0.339}& 0.152         & 0.238         &  & \textbf{0.525}& 0.401 & 0.335         &  \\
                          & 500& \textbf{0.377}& 0.263         & 0.237         &                        & \textbf{0.377}& 0.158         & 0.322         &  & \textbf{0.562}& 0.399 & 0.431         &  \\\midrule
\multirow[c]{3}{*}{\model}& 100& \textbf{0.452}& 0.223         & 0.261         & \multirow{3}{*}{-0.185}& \textbf{0.443}& 0.358         & 0.394         & \multirow{3}{*}{-0.430}& \textbf{0.529}& 0.404 & 0.418         & \multirow{3}{*}{-0.326} \\
                          & 300& \textbf{0.398}& 0.201         & 0.311         &                        & \textbf{0.494}& 0.392         & 0.353         & & \textbf{0.533}& 0.368 & 0.362         &  \\
                          & 500& \textbf{0.379}& 0.187         & 0.345         &                        & \textbf{0.505}& 0.356         & 0.344         & & \textbf{0.550}& 0.377 & 0.379         &  \\\midrule
\multirow[c]{3}{*}{\pos}  & 100& 0.188         & 0.089         & \textbf{0.196}& \multirow{3}{*}{-0.042}& \textbf{0.152}& 0.038         & 0.094         & \multirow{3}{*}{0.030} & \textbf{0.072}& -0.026& 0.024         & \multirow{3}{*}{-0.026} \\
                          & 300& 0.248         & 0.104         & \textbf{0.322}&                        & \textbf{0.197}& 0.010         & 0.189         &  & 0.116         & -0.015& \textbf{0.139}&  \\
                          & 500& 0.252         & 0.097         & \textbf{0.327}&                        & 0.203         & 0.013         & \textbf{0.213}&  & \textbf{0.143}& -0.026& 0.143         &  \\\midrule
\multirow[c]{3}{*}{\spell}& 100& 0.469         & 0.460         & \textbf{0.475}& \multirow{3}{*}{-0.103}& 0.401         & 0.441         & \textbf{0.483}& \multirow{3}{*}{0.109} & \textbf{0.278}& 0.235 & 0.235         & \multirow{3}{*}{0.032} \\
                          & 300& 0.427         & \textbf{0.453}& 0.434         &                        & 0.376         & 0.429         & \textbf{0.433}&  & 0.245         & 0.249 & \textbf{0.254}&  \\
                          & 500& 0.441         & \textbf{0.445}& 0.430         &                        & 0.384         & 0.438         & \textbf{0.453}&  & 0.237         & 0.249 & \textbf{0.254}&  \\\midrule
\multirow[c]{3}{*}{\syn}  & 100& 0.564         & \textbf{0.564}& 0.221         & \multirow{3}{*}{-0.071}& \textbf{0.223}& 0.134         & 0.034         & \multirow{3}{*}{-0.050}& \textbf{0.239}& 0.239 & 0.114         & \multirow{3}{*}{-0.041} \\
                          & 300& \textbf{0.627}& 0.574         & 0.525         &                        & \textbf{0.314}& 0.110         & 0.094         & & \textbf{0.363}& 0.332 & 0.296         &  \\
                          & 500& \textbf{0.696}& 0.590         & 0.633         &                        & \textbf{0.328}& 0.125         & 0.148         & & \textbf{0.391}& 0.332 & 0.321         &  \\\midrule
\multirow[c]{3}{*}{\topic}& 100& 0.181         & 0.162         & \textbf{0.208}& \multirow{3}{*}{0.098} & 0.192         & 0.131         & \textbf{0.206}& \multirow{3}{*}{0.142} & \textbf{0.152}& 0.062 & 0.096         & \multirow{3}{*}{0.107} \\
                          & 300& 0.247         & 0.139         & \textbf{0.349}&                        & 0.224         & 0.111         & \textbf{0.324}&  & 0.157         & 0.062 & \textbf{0.216}&  \\
                          & 500& 0.242         & 0.133         & \textbf{0.339}&                        & 0.251         & 0.114         & \textbf{0.345}&  & 0.170         & 0.062 & \textbf{0.231}&  \\\midrule
\multicolumn{2}{l|}{Best Conf. Count} &  \textbf{11} & 5 & 8 & 0 & \textbf{15} & 2 & 7 & 0 & \textbf{19} & 0 & 5 & 0  \\
\bottomrule
\end{tabular}
} %
\end{table}

Table~\ref{tab:compare_metrics} compares the correlation between actual pass rates from ten inferences and the predicted likelihood of passing obtained from \name to baselines. Scores derived from distributional deviation---\name, MDSA, and \gmmb---show weak to moderate correlations with observed pass rates, while SLL does not exhibit any meaningful correlation. Among the three distribution-based methods, \name achieves the highest correlation in 14 of the 24 configurations (8 tasks $\times$ 3 LLMs, each with a reference set size of 500), while MDSA and \gmmb perform best in only 3 and 7 configurations, respectively.

\begin{figure}[ht]
    \centering
    \includegraphics[width=0.65\linewidth]{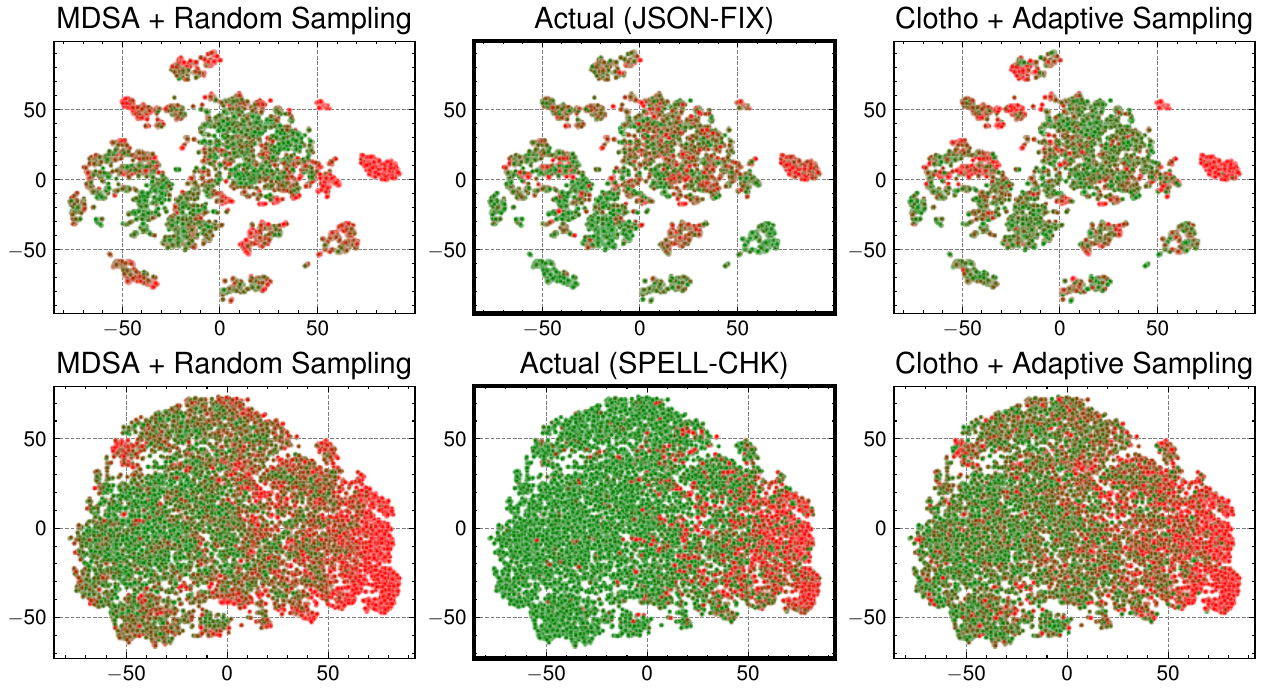}
    \caption{Projection of prediction scores from \name, MDSA, and GT pass rates (Llama, $N$=500).}
    \label{fig:rq1-1/clotho_tsne_examples}
\end{figure}

We observe that \name outperforms MDSA on some tasks (e.g., \json), while the opposite happens for other tasks (e.g., \spell). We posit that this can be explained by the structures in input distribution. Figure~\ref{fig:rq1-1/clotho_tsne_examples} shows the latent space of LIHS for all inputs of \json and \spell, each with predicted likelihood of passing from MDSA (left) and \name (right), as well as the actual ground truth pass rates (centre). We observe that the input distribution of \json task has complex structures with multiple clusters, and the unimodal distribution used by MDSA cannot capture the likelihood of passing over the complex distribution. For example, compare the pass and fail prediction for the bottom left cluster of \json: \name predicts most of the inputs to pass (which aligns with the ground truths), whereas MDSA predicts most of them to be incorrect primarily because they are far from the mean of the main unimodal distribution. In contrast, the input distribution of \spell task forms a large single cluster, where the centres of the passing and failing inputs are found in the left and the right part of the cluster, respectively. In cases like this, the distance from the mean of the unimodal distribution may be a reasonably accurate predictor of passing.

Similarly, \gmmb outperforms \name in three tasks: \spell, \gh, and \topic. Note that \gmmb has two major differences from \name: it samples inputs into the reference set randomly, and its GMM does not update component number and representation dimensionality adaptively. We note that \gh exhibits a distribution close to unimodal, similar to that of \spell. The structure of input distribution for \topic is more complex, but it is more contiguous when compared to tasks like \json. Our conjecture is that the smaller number of mixture components used by \gmmb (which is the initial value, five), as well as random sampling of inputs into the reference set, may perform reasonably well for tasks like these. Overall, the observed differences support the use of balanced sampling and adaptive modelling used by \name. 

\subsubsection{RQ1-2: Comparison of Different Sampling Strategies}

\begin{figure}[ht]
    \centering
    \includegraphics[width=\linewidth]{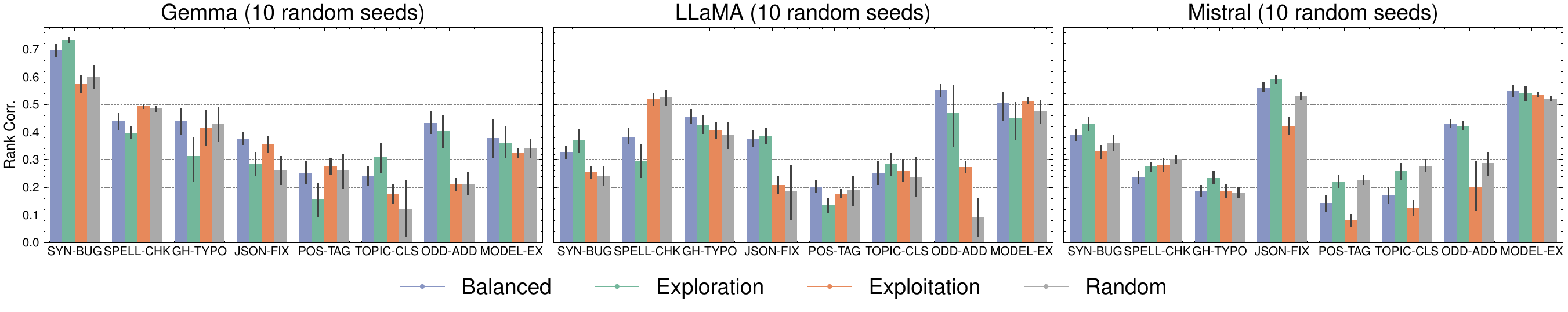}
    \caption{Spearman Rank Correlations from Different Sampling Strategies ($N$=500)}
    \label{fig:rq1_2/comparison_samplings_n500}
\end{figure}

We now consider the impact of sampling strategy in more detail. Figure~\ref{fig:rq1_2/comparison_samplings_n500} compares the correlation between predicted likelihood of passing and actual pass rates, obtained with different sampling strategies, when we build a reference set of 500 inputs. \name's balanced sampling strategy shows the best average performance, 16.06\% higher correlation than random and 2.54\% higher than exploration-based sampling (the second best). However, it also shows that, depending on tasks, the best sampling strategy can differ: on Gemma and Llama, balanced sampling outperforms exploration by 10.05\% and 8.05\%, respectively, while on Mistral, exploration surpasses balanced by 10.17\%. We analyse the input distribution in more detail to explain these differences.

\begin{figure}[ht]
    \centering
    \includegraphics[width=\linewidth]{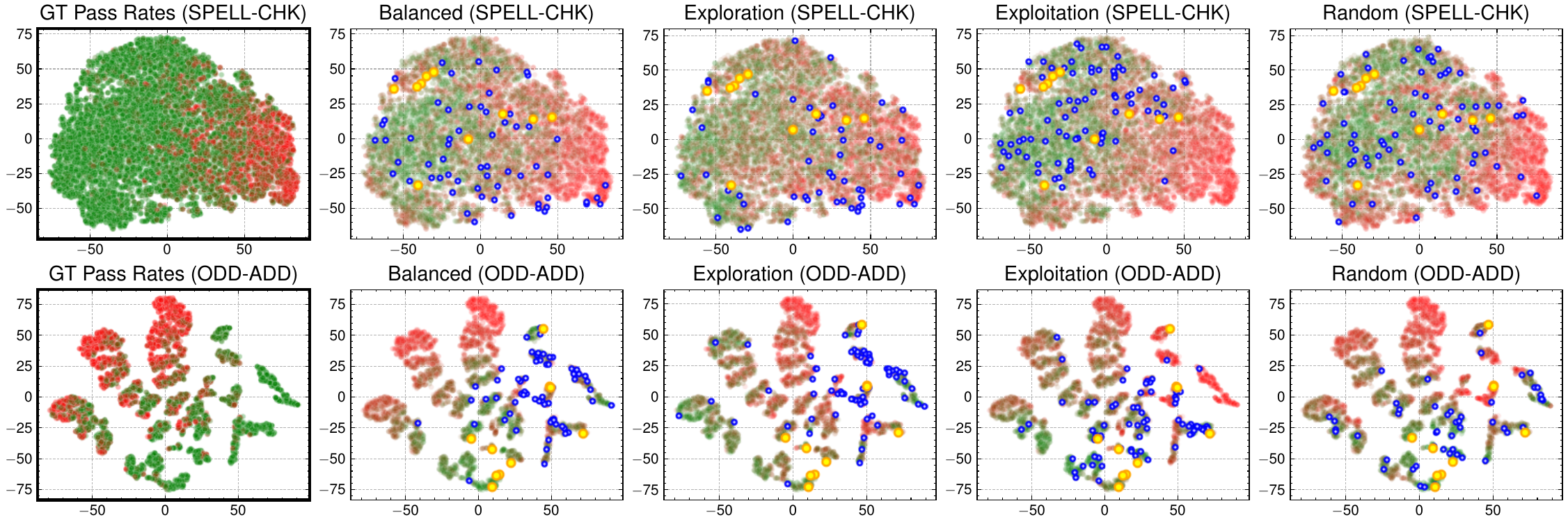}
    \caption{Distribution of selected reference points with each sampling strategy (Llama, $N$=100)}
    \label{fig:rq1_2/reference_points_visualization}
\end{figure}

Figure~\ref{fig:rq1_2/reference_points_visualization} shows 100 reference inputs for \spell and \texttt{ODD-ADD}, sampled by different sampling strategies from Llama, denoted in blue. The yellow points, on the other hands, are the ten inputs in the initial reference set. Each plot additionally shows the resulting \name scores for all the remaining inputs in red (likely to fail) and green (likely to pass), which can be compared to the ground truth pass rate distribution shown in GT Pass Rate (leftmost). Inputs for \spell form a single contiguous cluster with two polar ends, whereas inputs for \odd form multiple, fragmented clusters. We note that exploration-based sampling for \spell tends to prefer inputs that are outliers, i.e., far from the already sampled reference set. The shape of distribution of \spell means that such outliers are more likely to fail, making little contribution to the modelling of the passing input distribution. In contrast, exploitation-based sampling achieves a better coverage of passing inputs, as it tries to clarify the boundaries between mixture components fitted to the initially given passing inputs. Random sampling, on the other hand, achieves a good coverage of the entire input space. Both exploitation-based and random sampling result in more accurate modelling of the input distribution, leading to higher correlation than exploration-based sampling. In contrast, \odd has fragmented and scattered clusters. Exploration-based sampling successfully discovers outer clusters and accurately models their distributions, whereas exploitation-based and random sampling fail to do so, resulting in lower correlations.

A closer examination of Figure~\ref{fig:rq1_2/comparison_samplings_n500} also reveals that Mistral shows a different pattern from those of Llama and Gemma: exploitation-based sampling performs poorly in general, whereas random sampling performs better than for Llama and Gemma. We conjecture that this is because Mistral exhibits substantially higher failure rates across all tasks. For example, in \json task, only 31\% of inputs achieve pass rates above 0.5, compared to 67\% for Llama and 80\% for Gemma. Such lower pass rate results in a larger number of smaller, more fragmented clusters of passing inputs. Under limited initial data, the GMM may struggle to capture the global structure, and its entropy-based signal can become less reliable, leading to sampling that is biased toward local regions.

\begin{rqbox}[Answer to RQ1]
\name successfully predicts pass/fail likelihoods of inputs before the LLM generates outputs, showing clearer advantages over baselines on complex, multi-cluster input distributions. Sampling strategy significantly impacts performance; across tasks, balanced sampling yields the most stable results.
\end{rqbox}

\subsection{Test Prioritisation Effectiveness (RQ2)}

RQ2 concerns how well \name can prioritise inputs that are likely to fail for more efficient testing. Table~\ref{tab:rq3_auc_stat_test} shows the ROC-AUC of adequacy scores from \name and MDSA for the prediction of failure, as well as the $p$-values from Mann-Whitney U test of the adequacy scores between passing and failing inputs. \name achieves a ROC-AUC of 0.716 for unseen inputs across all three models (i.e., those not included in the reference set), when using 500 labelled references (which are, on average, 5.4\% of inputs for tasks). In contrast, MDSA achieves a ROC-AUC of 0.655. For all tasks and models, both \name and MDSA produce adequacy scores that can distinguish passing and failing inputs with statistical significance.

\begin{table}[h!]
    \centering
    \caption{ROC-AUC scores and $p$-value from Mann-Whitney U Test pass/fail prediction ($N=500$).}
    \label{tab:rq3_auc_stat_test}
    \resizebox{\textwidth}{!}{%
\begin{tabular}{lll|ll|ll|ll|ll|ll}
\toprule
 & \multicolumn{6}{c}{ROC-AUC} & \multicolumn{6}{c}{$p$-value (Mann-Whitney)} \\
\midrule
 & \multicolumn{2}{c}{Gemma 2 9B} & \multicolumn{2}{c}{Llama 3.1 8B} & \multicolumn{2}{c}{Mistral 7B} & \multicolumn{2}{c}{Gemma 2 9B} & \multicolumn{2}{c}{Llama 3.1 8B} & \multicolumn{2}{c}{Mistral 7B} \\
\midrule
\multicolumn{1}{l|}{Task ID} & \name & MDSA & \name & MDSA & \name & MDSA & \name & MDSA & \name & MDSA & \name & MDSA \\
\midrule
\multicolumn{1}{l|}{\odd} & \textbf{0.714} $\pm$ 0.04 & 0.609 $\pm$ 0.01 & \textbf{0.790} $\pm$ 0.02 & 0.609 $\pm$ 0.02 & \textbf{0.883} $\pm$ 0.01 & 0.823 $\pm$ 0.03 & 2.4e-81 & 4.1e-34 & 6.1e-254 & 2.3e-17 & 6.3e-117 & 6.5e-84 \\
\multicolumn{1}{l|}{\gh} & 0.737 $\pm$ 0.05 & \textbf{0.756} $\pm$ 0.02 & 0.731 $\pm$ 0.02 & \textbf{0.754} $\pm$ 0.01 & \textbf{0.587} $\pm$ 0.02 & 0.575 $\pm$ 0.01 & 9.3e-105 & 1.2e-293 & 5.6e-238 & 0.0e+00 & 2.4e-11 & 2.2e-23 \\
\multicolumn{1}{l|}{\json} & \textbf{0.764} $\pm$ 0.02 & 0.642 $\pm$ 0.01 & \textbf{0.690} $\pm$ 0.02 & 0.583 $\pm$ 0.01 & \textbf{0.793} $\pm$ 0.01 & 0.702 $\pm$ 0.02 & 4.6e-122 & 7.0e-45 & 1.8e-73 & 8.0e-20 & 1.4e-253 & 7.7e-84 \\
\multicolumn{1}{l|}{\model} & \textbf{0.696} $\pm$ 0.07 & 0.595 $\pm$ 0.02 & \textbf{0.758} $\pm$ 0.05 & 0.684 $\pm$ 0.02 & \textbf{0.788} $\pm$ 0.02 & 0.702 $\pm$ 0.02 & 2.6e-36 & 2.8e-24 & 6.3e-104 & 4.0e-141 & 0.0e+00 & 3.0e-183 \\
\multicolumn{1}{l|}{\pos} & \textbf{0.648} $\pm$ 0.04 & 0.568 $\pm$ 0.01 & \textbf{0.601} $\pm$ 0.02 & 0.516 $\pm$ 0.01 & \textbf{0.571} $\pm$ 0.02 & 0.508 $\pm$ 0.01 & 4.4e-41 & 1.4e-17 & 2.6e-45 & 1.8e-02 & 1.7e-09 & 2.1e-01 \\
\multicolumn{1}{l|}{\spell} & 0.869 $\pm$ 0.04 & \textbf{0.871} $\pm$ 0.01 & 0.748 $\pm$ 0.03 & \textbf{0.776} $\pm$ 0.01 & 0.649 $\pm$ 0.02 & \textbf{0.663} $\pm$ 0.01 & 2.6e-186 & 0.0e+00 & 2.4e-182 & 7.0e-266 & 1.9e-52 & 1.2e-96 \\
\multicolumn{1}{l|}{\syn} & \textbf{0.905} $\pm$ 0.02 & 0.836 $\pm$ 0.01 & \textbf{0.714} $\pm$ 0.02 & 0.593 $\pm$ 0.01 & \textbf{0.717} $\pm$ 0.01 & 0.679 $\pm$ 0.01 & 0.0e+00 & 0.0e+00 & 0.0e+00 & 1.5e-55 & 0.0e+00 & 4.8e-284 \\
\multicolumn{1}{l|}{\topic} & \textbf{0.640} $\pm$ 0.04 & 0.588 $\pm$ 0.01 & \textbf{0.618} $\pm$ 0.04 & 0.557 $\pm$ 0.02 & \textbf{0.580} $\pm$ 0.03 & 0.539 $\pm$ 0.01 & 1.9e-29 & 3.2e-19 & 4.8e-10 & 1.7e-06 & 2.0e-06 & 3.7e-04 \\ \midrule
\multicolumn{1}{l|}{Average} & \textbf{0.746} $\pm$ 0.10 & 0.683 $\pm$ 0.11& \textbf{0.706} $\pm$ 0.07 & 0.634 $\pm$ 0.09 & \textbf{0.696} $\pm$ 0.11  & 0.649 $\pm$ 0.10  & \multicolumn{6}{c}{-} \\
\bottomrule
\end{tabular}
} %
\end{table}

\begin{table}[ht]
\caption{Comparison of \fan for all tasks and models ($\mu$: mean, $\sigma$: standard deviation).}
\label{tab:rq2_fail_at_N}
\resizebox{\textwidth}{!}{%
\begin{tabular}{lr|rrrrrr|rrrrrr|rrrrrr}
\toprule
\multicolumn{2}{c}{} & \multicolumn{6}{c}{Gemma 2 9B} & \multicolumn{6}{c}{Llama 3.1 8B} & \multicolumn{6}{c}{Mistral 7B} \\ \midrule
 \multicolumn{2}{c}{} & \multicolumn{2}{c}{\name} & \multicolumn{2}{c}{MDSA} & \multicolumn{2}{c}{Random} & \multicolumn{2}{c}{\name} & \multicolumn{2}{c}{MDSA} & \multicolumn{2}{c}{Random} & \multicolumn{2}{c}{\name} & \multicolumn{2}{c}{MDSA} & \multicolumn{2}{c}{Random} \\ \midrule
Task ID & $N$ & $\mu$ & $\sigma$ & $\mu$ & $\sigma$ & $\mu$ & $\sigma$ & $\mu$ & $\sigma$ & $\mu$ & $\sigma$ & $\mu$ & $\sigma$ & $\mu$ & $\sigma$ & $\mu$ & $\sigma$ & $\mu$ & $\sigma$ \\
\midrule
\multirow[c]{3}{*}{\odd}   & 100 & \textbf{100.0}& 0.00 & \textbf{100.0}& 0.00 & 40.7 & 4.06 & \textbf{95.9} & 7.22 & 29.0          & 18.62& 38.1 & 3.11 & \textbf{99.9} & 0.32 & 96.4          & 4.43 & 91.4 & 1.90 \\
                           & 300 &  299.5         & 0.71 & \textbf{300.0}& 0.00 & 124.7& 9.08 & \textbf{284.0}& 11.98& 123.9         & 59.64& 117.2& 5.53 & \textbf{299.7}& 0.95 & 293.5         & 6.36 & 272.3& 4.79 \\
                           & 500 &  455.4         & 13.31& \textbf{477.3}& 6.72 & 200.8& 9.27 & \textbf{459.9}& 14.39& 219.1         & 77.32& 198.8& 10.59& \textbf{499.4}& 1.58 & 492.7         & 7.06 & 455.4& 6.22 \\ \cmidrule{1-20}
\multirow[c]{3}{*}{\gh}    & 100 &  \textbf{86.7} & 5.01 & 77.5          & 7.75 & 34.6 & 3.86 & \textbf{95.7} & 2.06 & 82.4          & 2.80 & 48.4 & 4.14 & \textbf{92.0} & 3.20 & 83.7          & 2.79 & 69.5 & 4.79 \\
                           & 300 &  \textbf{247.6}& 17.15& 246.3         & 15.37& 106.8& 5.18 & \textbf{278.0}& 11.10& 251.4         & 10.34& 146.2& 9.45 & \textbf{259.3}& 18.54& 247.9         & 5.38 & 206.1& 12.84 \\
                           & 500 &  402.2         & 33.27& \textbf{408.3}& 23.98& 179.9& 6.06 & \textbf{453.2}& 18.28& 429.7         & 13.54& 244.3& 13.43& \textbf{417.4}& 26.53& 404.0         & 11.12& 342.0& 12.63 \\\cmidrule{1-20}
\multirow[c]{3}{*}{\json}  & 100 &  \textbf{76.7} & 7.09 & 67.5          & 9.03 & 16.2 & 3.77 & \textbf{92.3} & 4.00 & 89.8          & 0.79 & 24.9 & 5.30 & 99.9          & 0.32 & \textbf{100.0}& 0.00 & 58.8 & 7.02 \\
                           & 300 &  \textbf{220.7}& 24.33& 133.9         & 7.99 & 51.4 & 5.76 & 230.3         & 8.62 & \textbf{234.3}& 8.17 & 75.6 & 6.40 & \textbf{296.9}& 1.20 & 295.6         & 0.52 & 177.0& 11.82 \\
                           & 500 &  \textbf{326.3}& 39.50& 195.0         & 13.68& 85.9 & 4.07 & \textbf{316.7}& 10.14& 305.6         & 11.04& 125.5& 10.56& \textbf{478.3}& 4.52 & 454.7         & 16.50& 299.0& 10.67 \\\cmidrule{1-20}
\multirow[c]{3}{*}{\model} & 100 &  \textbf{45.3} & 17.64& 27.0          & 2.45 & 33.9 & 3.84 & \textbf{69.7} & 6.04 & 39.0          & 7.15 & 44.0 & 4.71 & \textbf{75.7} & 5.50 & 41.0          & 5.85 & 46.5 & 3.69 \\
                           & 300 &  \textbf{143.2}& 41.57& 88.0          & 7.21 & 101.6& 7.18 & \textbf{205.2}& 16.39& 132.1         & 8.05 & 134.2& 9.92 & \textbf{231.3}& 10.65& 146.4         & 13.55& 144.6& 9.36 \\
                           & 500 &  \textbf{240.0}& 65.77& 162.9         & 13.80& 167.1& 10.65& \textbf{337.5}& 19.85& 245.7         & 10.73& 224.7& 13.86& \textbf{383.5}& 12.96& 278.1         & 17.93& 245.5& 12.31 \\\cmidrule{1-20}
\multirow[c]{3}{*}{\pos}   & 100 &  76.0          & 13.60& \textbf{78.2} & 1.75 & 22.3 & 3.80 & 64.3          & 7.93 & \textbf{74.6} & 1.26 & 25.1 & 4.77 & 86.6          & 4.90 & \textbf{91.5} & 0.53 & 28.0 & 3.43 \\
                           & 300 &  \textbf{199.6}& 39.97& 190.2         & 12.17& 59.1 & 4.12 & 154.0         & 13.17& \textbf{166.7}& 9.38 & 69.3 & 7.41 & 218.4         & 12.96& \textbf{224.0}& 2.00 & 87.2 & 7.91 \\
                           & 500 &  282.6         & 53.80& \textbf{283.4}& 19.02& 99.7 & 5.76 & \textbf{239.3}& 24.53& 228.4         & 8.66 & 116.7& 9.82 & \textbf{325.6}& 21.88& 320.4         & 8.75 & 148.7& 8.84 \\\cmidrule{1-20}
\multirow[c]{3}{*}{\spell} & 100 &  \textbf{86.9} & 2.60 & 82.0          & 0.82 & 12.3 & 4.32 & \textbf{75.3} & 6.17 & 69.2          & 1.55 & 18.3 & 3.68 & \textbf{70.9} & 5.04 & 60.1          & 3.00 & 21.7 & 3.20 \\
                           & 300 &  \textbf{238.9}& 6.08 & 223.0         & 12.25& 37.5 & 7.14 & \textbf{208.8}& 19.01& 207.7         & 10.49& 56.3 & 5.29 & \textbf{199.5}& 16.03& 170.8         & 15.63& 61.6 & 6.59 \\
                           & 500 &  \textbf{362.2}& 14.73& 331.8         & 19.56& 61.8 & 7.04 & 313.4         & 31.92& \textbf{335.2}& 23.08& 92.9 & 11.10& \textbf{301.4}& 19.78& 264.8         & 19.53& 104.5& 6.60 \\\cmidrule{1-20}
\multirow[c]{3}{*}{\syn}   & 100 &  94.4          & 3.17 & \textbf{96.2} & 2.86 & 58.9 & 4.68 & \textbf{84.8} & 5.49 & 76.7          & 3.56 & 60.6 & 4.67 & 87.4          & 4.95 & \textbf{88.1} & 4.75 & 69.4 & 4.62 \\
                           & 300 &  285.1         & 6.67 & \textbf{287.8}& 7.84 & 173.0& 8.25 & \textbf{254.1}& 11.24& 231.7         & 7.73 & 188.6& 8.71 & \textbf{267.5}& 8.06 & 261.8         & 12.66& 206.3& 7.39 \\
                           & 500 &  477.7         & 8.60 & \textbf{480.5}& 9.78 & 286.6& 8.66 & \textbf{420.1}& 15.07& 383.6         & 14.84& 312.0& 9.10 & \textbf{448.9}& 11.01& 438.9         & 16.31& 341.0& 7.90 \\\cmidrule{1-20}
\multirow[c]{3}{*}{\topic} & 100 &  52.0          & 10.41& \textbf{59.9} & 3.51 & 15.5 & 3.31 & 52.0          & 9.73 & \textbf{64.9} & 3.35 & 20.2 & 3.19 & 49.8          & 6.66 & \textbf{62.2} & 2.04 & 25.7 & 3.71 \\
                           & 300 &  129.3         & 22.70& \textbf{154.0}& 8.38 & 50.8 & 5.73 & 123.1         & 16.06& \textbf{154.7}& 7.21 & 62.6 & 5.74 & 121.6         & 14.51& \textbf{129.4}& 5.10 & 70.4 & 8.10 \\
                           & 500 &  193.6         & 29.28& \textbf{217.2}& 15.52& 88.4 & 11.21& 185.6         & 26.62& \textbf{223.6}& 9.61 & 107.2& 7.15 & \textbf{186.1}& 15.32& 177.4         & 6.11 & 115.6& 8.80 \\
\bottomrule
\end{tabular}
} %
\end{table}

Table~\ref{tab:rq2_fail_at_N} shows \fan, i.e., the number of failure-inducing inputs (i.e., those whose pass rates are 0.5 or below) out of $N$ inputs selected in the descending order of adequacy metrics ($N=\{100, 300, 500\}$): for each pair of metric and $N$, we report average and standard deviation of \fan from 10 different orderings, each obtained from repeated runs. Note that the column Random represents random ordering and, therefore, the corresponding \fan values reflect the average ratio of failure-inducing inputs in the entire dataset for each task. The best \fan for each configuration is typeset in bold. \name shows strong capability in prioritising failure-inducing inputs, with 80.4\% of the top-100 ranked inputs and 75.2\% of top-300 ranked inputs exposing actual failures, compared to 73.1\% and 68.2\% for MDSA, respectively, when aggregated over all tasks.
Notably, \name can significantly outperform MDSA on tasks where MDSA struggles (e.g., \odd or \model). When MDSA outperforms \name, the margins tend to be smaller. These results highlight the practical value of \name: by ranking inputs, it helps testers focus execution and labelling effort on the most challenging cases.

\begin{rqbox}[Answer to RQ2]
\name's learnt adequacy scores can reliably identify failure-inducing inputs on unseen data. For test prioritisation, \name consistently ranks a high proportion of failures in the top positions, providing evidence that it can help testers focus labelling effort on the most challenging inputs.
\end{rqbox}

\subsection{RQ3: Complementarity Analysis}

\begin{table}[ht]
\caption{Comparison between \name and post-generation uncertainty metrics across all models and tasks.}
\label{tab:loh_metrics}
\resizebox{\textwidth}{!}{%
\begin{tabular}{l|ccccc|ccccc|ccccc}
\toprule
\multicolumn{1}{c}{} & \multicolumn{5}{c}{Gemma 2 9B} & \multicolumn{5}{c}{Llama 3.1 8B} & \multicolumn{5}{c}{Mistral 7B} \\
\midrule
Task ID & \lohsv & \sem & \tokp & \toke & \name & \lohsv & \sem & \tokp & \toke & \name & \lohsv & \sem & \tokp & \toke & \name \\
\midrule
\odd & 0.390 & \textbf{0.487} & -0.138 & \underline{0.458} & 0.434 & 0.519 & \textbf{0.730} & 0.548 & 0.467 & \underline{0.551} & -0.076 & \underline{0.282} & 0.004 & 0.005 & \textbf{0.432} \\
\gh & 0.534 & \textbf{0.558} & 0.284 & \underline{0.540} & 0.440 & \textbf{0.800} & \underline{0.543} & 0.382 & 0.465 & 0.456 & \underline{0.614} & \textbf{0.633} & 0.325 & 0.430 & 0.188 \\
\json & \underline{0.342} & 0.308 & 0.138 & 0.230 & \textbf{0.377} & \textbf{0.688} & \underline{0.442} & 0.277 & 0.319 & 0.377 & \textbf{0.632} & 0.403 & 0.226 & 0.246 & \underline{0.562} \\
\model & \underline{0.348} & 0.124 & -0.184 & 0.177 & \textbf{0.379} & \textbf{0.650} & 0.223 & 0.173 & 0.196 & \underline{0.505} & \textbf{0.581} & 0.237 & 0.062 & 0.051 & \underline{0.550} \\
\pos & \underline{0.310} & \textbf{0.313} & 0.026 & 0.218 & 0.252 & \textbf{0.726} & \underline{0.367} & 0.274 & 0.321 & 0.203 & \textbf{0.693} & \underline{0.410} & 0.221 & 0.263 & 0.143 \\
\spell & \textbf{0.461} & 0.425 & 0.263 & 0.415 & \underline{0.441} & \textbf{0.722} & \underline{0.526} & 0.430 & 0.480 & 0.384 & \textbf{0.520} & \underline{0.495} & 0.233 & 0.279 & 0.237 \\
\syn & -0.069 & 0.090 & 0.007 & 0.055 & \textbf{0.696} & 0.161 & 0.194 & 0.188 & \underline{0.219} & \textbf{0.328} & -0.168 & -0.092 & -0.040 & -0.036 & \textbf{0.391} \\
\topic & \textbf{0.364} & 0.174 & 0.039 & 0.121 & \underline{0.242} & \textbf{0.556} & 0.175 & 0.142 & 0.152 & \underline{0.251} & \textbf{0.597} & \underline{0.234} & 0.139 & 0.180 & 0.170 \\
\bottomrule
\end{tabular}
} %
\end{table}

RQ3 compares \name to post-generation approaches. Table~\ref{tab:loh_metrics} begins by comparing the results of correlation analysis, i.e., how well the adequacy scores correlate with pass rates, across \name and post-generation metrics: the highest are typeset in bold, and the second highest underlined. Post-generation variance-based metrics generally perform well, with \lohsv achieving the strongest correlations overall (6/8 for Llama, 5/8 for Mistral). In contrast, token-based confidence measures generally show only moderate correlations across all configurations. Overall, \name demonstrates strong performance, despite lacking access to the actual model outputs. Notably, \name achieves the best performance on specific tasks, particularly for \syn, where post-generation approaches typically underperform. We also note that the performance distribution varies significantly across model architectures. For Gemma, \name achieves comparable performance to variance-based measures, suggesting particular compatibility with certain model families.

Importantly, these results highlight that \name can achieve competitive performance without requiring full output generation, leading to significant efficiency gains. Extracting LIHS incurs negligible additional cost compared to standard inference, taking only a few minutes versus several days for full generation across a test suite. The training cost of the GMM is also minimal, with fitting on 500 samples taking less than 14 seconds (4.4 seconds on average), including time for hyperparameter adaptation.

\begin{figure}[t]
    \centering
    \includegraphics[width=\linewidth]{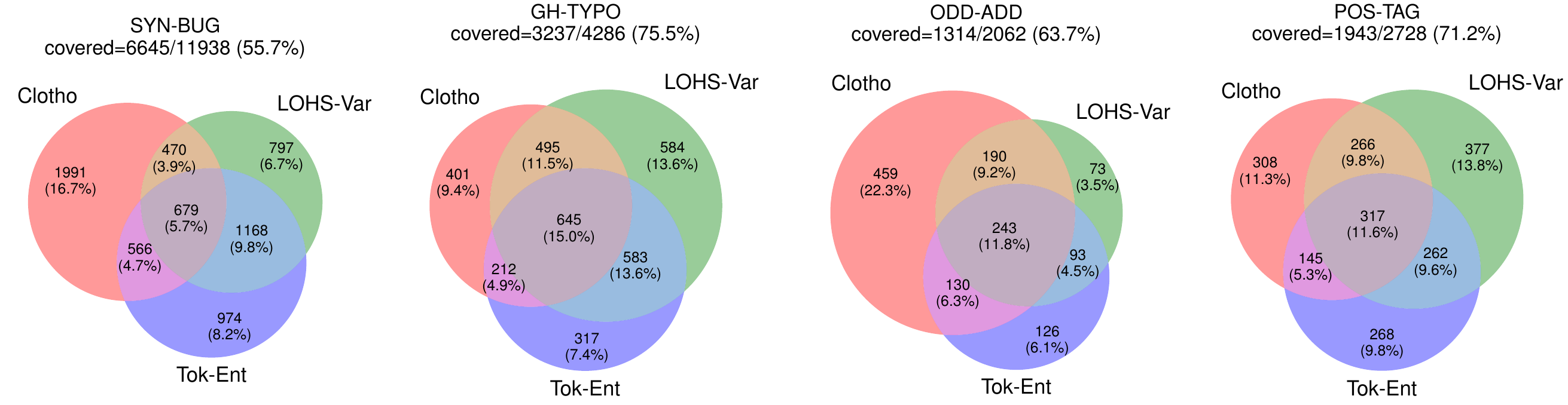}
    \caption{3-way Venn diagrams drawn from samples with high adequacy scores (top 25\%) each from \name, \lohsv, and \toke among high-failure-rate inputs ($ p_f \ge 0.8$) in Llama.}
    \label{fig:rq3_1/3_way_complemantary_venn_diagram}
\end{figure}

Given the variance in correlations, we investigate whether these metrics can complement each other in order to detect inputs with high failure rates. First, we choose inputs that fail eight or more times out of 10 repeated runs. Subsequently, we compute three representative adequacy metrics: \name, \lohsv, \toke, and choose the inputs with the top 25\% scores. Figure~\ref{fig:rq3_1/3_way_complemantary_venn_diagram} shows Venn diagram of such high adequacy, high failure rate inputs for four tasks: \syn, \gh, \odd, and \pos. The number of covered inputs shows how many of the high failure rate inputs are also considered as high adequacy inputs by the three metrics analysed: collectively, at least 55.7\% (\syn) or more of those inputs show top 25\% adequacy scores by at least one of the metrics. The diagrams also show the complementary relationship between the metrics: unanimous agreements between three metrics range only from 5.7\% to 15\% of high failure rate inputs, while individual metrics uniquely account for up to 22.3\% of them (\name for \odd). This suggests that combining the metrics can broaden coverage and capture a larger portion of failure-inducing inputs, indicating that although \name was originally designed as a pre-generation metric, it can also provide complementary benefits when integrated with post-generation metrics to better focus labelling efforts.

\begin{rqbox}[Answer to RQ3]
\name's failure prediction performance is comparable to those of post-generation metrics, despite not having access to the actual generated outputs. Closer analysis shows they are complementary, each capturing distinct subsets of failure-inducing inputs, indicating the potential of integrating \name with post-generation metrics.
\end{rqbox}

\begin{table}[ht]
\centering
\caption{Spearman's rank correlation between scores obtained from OLMs with reference set size 500 and actual pass rates of proprietary LLMs.}
\label{tab:olm_adequacy_transfer}
\resizebox{0.8\textwidth}{!}{%
\begin{tabular}{lllll|llll|llll}
\toprule
 & \multicolumn{4}{c}{Gemma 2 9B} & \multicolumn{4}{c}{Llama 3.1 8B} & \multicolumn{4}{c}{Mistral 7B} \\
\midrule
\multicolumn{1}{l|}{Task ID} & SELF & Claude & GPT & Gemini & SELF & Claude & GPT & Gemini & SELF & Claude & GPT & Gemini \\
\midrule
\multicolumn{1}{l|}{\odd}  & 0.444& \underbar{\textbf{0.447}}& 0.094         & 0.064         & 0.549& 0.411         & \textbf{0.439}& 0.235         & 0.434& 0.231         & \underbar{\textbf{0.437}}& 0.182 \\
\multicolumn{1}{l|}{\gh}   & 0.438& 0.383         & 0.373         & \underbar{\textbf{0.442}}& 0.470& 0.380         & 0.365         & \textbf{0.413}& 0.197& 0.106         & \textbf{0.140}& 0.125 \\
\multicolumn{1}{l|}{\json} & 0.394& \textbf{0.128}& 0.108         & 0.111         & 0.360& 0.098         & \textbf{0.167}& 0.160         & 0.568& \textbf{0.203}& -0.010        & 0.006 \\
\multicolumn{1}{l|}{\model}& 0.386& 0.248         & 0.347         & \underbar{\textbf{0.402}}& 0.512& 0.248         & 0.341         & \textbf{0.427}& 0.555& 0.288         & 0.380         & \textbf{0.456} \\
\multicolumn{1}{l|}{\pos}  & 0.289& 0.267         & \textbf{0.272}& 0.253         & 0.221& 0.210         & \underbar{\textbf{0.235}}& 0.184         & 0.138& 0.190         & \underbar{\textbf{0.233}}& 0.173 \\
\multicolumn{1}{l|}{\spell}& 0.452& \underbar{\textbf{0.465}}& 0.443         & 0.419         & 0.374& \textbf{0.349}& 0.340         & 0.321         & 0.247& \underbar{\textbf{0.292}}& 0.263         & 0.239 \\
\multicolumn{1}{l|}{\syn}  & 0.707& 0.435         & \textbf{0.543}& 0.532         & 0.345& 0.190         & 0.178         & \textbf{0.224}& 0.396& 0.325         & \textbf{0.379}& 0.349 \\
\multicolumn{1}{l|}{\topic}& 0.252& 0.215         & \underbar{\textbf{0.259}}& 0.229         & 0.256& 0.188         & 0.201         & \textbf{0.209}& 0.181& 0.139         & 0.160         & \textbf{0.161} \\
\bottomrule
\end{tabular}
} %
\end{table}

\subsection{RQ4: Transferability to Proprietary LLMs}

\begin{figure}[t]
    \centering
    \includegraphics[width=\linewidth]{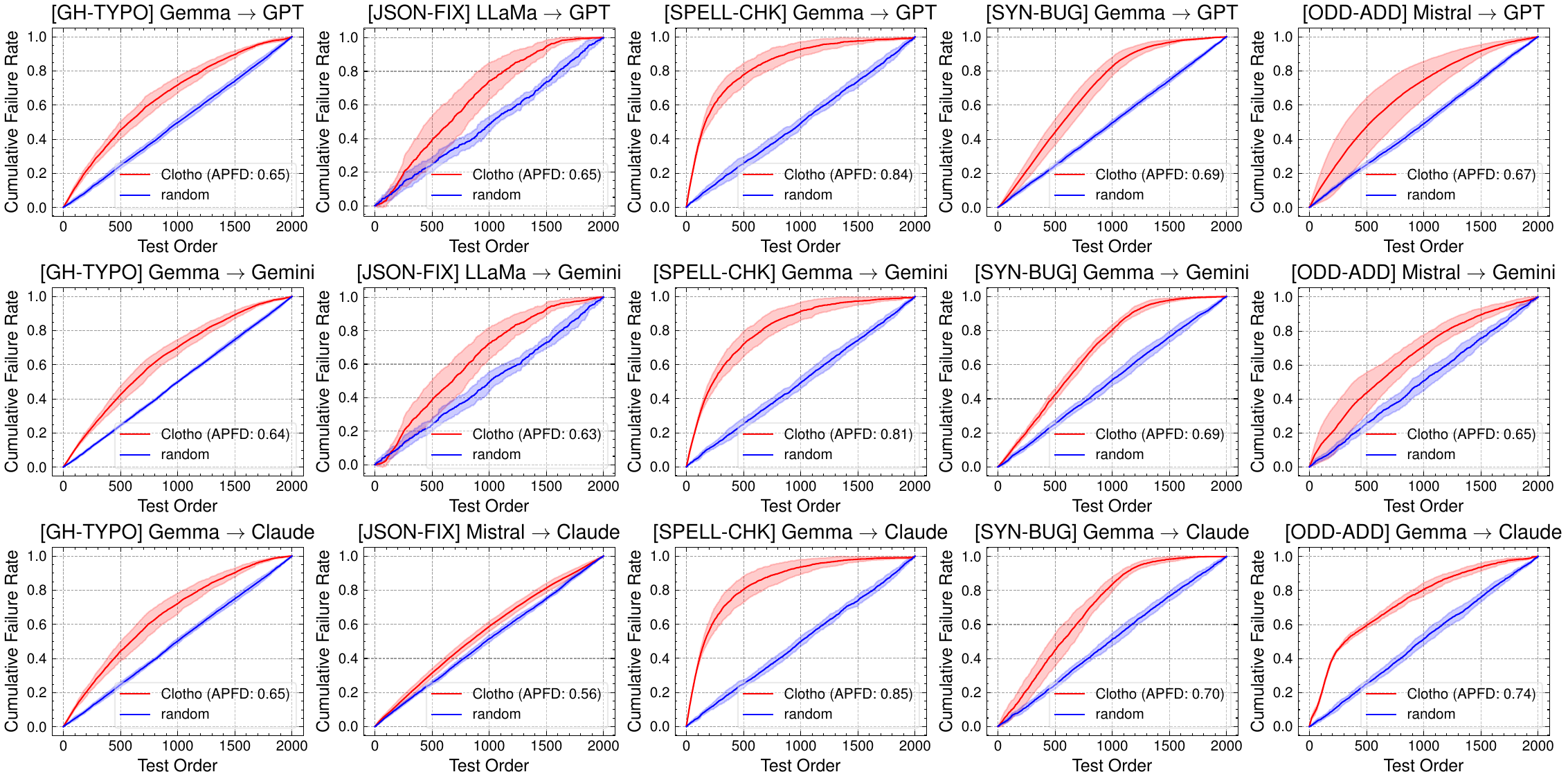}
    \caption{Average percentage of detected failures when scores from Open-weight Language Models (OLMs) are applied to proprietary LLMs.}
    \label{fig:rq4_apfd_proprietary}
\end{figure}

The final research question focuses on the more practical applicability of \name: by design, \name requires internal representations of the target LLMs that are only available from Open-weight Language Models (OLMs). However, some application may have to rely on stronger, proprietary closed-weight models. Table~\ref{tab:olm_adequacy_transfer} reports score correlation analysis results, in which the adequacy scores are computed by \name using OLMs, while the pass rates are obtained from 2,000 queries made to each of Claude (haiku), GPT (4o-mini), and Gemini (2.5-flash-lite). The results show that the adequacy scores predicted using OLMs can still produce moderate correlations with the actual pass rates of proprietary models: in some cases, they even outperform correlations with pass rates from the OLM itself (contained in the SELF column). For example, Gemma scores achieve higher correlation when applied to Gemini for \model than when applied to itself; similarly, Mistral scores perform better when applied to GPT than itself for \pos task. This suggests that \name can capture certain generalisable aspects of input difficulty that transcend specific model architectures.

We subsequently evaluate the effectiveness of these transferred adequacy scores for prioritising failure-inducing inputs for the proprietary models. Figure~\ref{fig:rq4_apfd_proprietary} shows the cumulative failure rates for the transferred input orderings for \gh, \json, \spell, \syn, and \odd for a subset of transfer pairs (best transfer pair for each task): the solid lines denote the average from 10 runs of \name, and the coloured bands denote the standard deviation. We also report the Average Percentage of Fault Detected (APFD)~\cite{Elbaum:2000rr}, which is essentially the area under curve. Given that average APFD would converge to 0.5 as the number of runs increases, the transferred ordering can achieve significantly effective test prioritisation. This is because proprietary models tend to have lower failure rate. As such, prioritising failure-inducing inputs early on can have bigger impact in proprietary models than in OLMs.
Table~\ref{tab:rq4_fail_at_N_proprietary} contains the \fan from transferred \name scores at $N$=500, as well as random orderings, per task: the large gains over random ordering aligns with high APFDs we observe in Figure~\ref{fig:rq4_apfd_proprietary}. Aggregated across tasks and models, \name finds on average 23.75 additional failures at $N$=100 and 61.15 at $N$=500 over random selection, corresponding to 126.8\% and 64.5\% improvements, respectively.

\begin{table}[t]
\centering
\caption{Number of detected failing inputs at $N$ labelling budgets (\fan) when scores from OLMs are applied to proprietary model ($N$=500).}
\label{tab:rq4_fail_at_N_proprietary}
\resizebox{\textwidth}{!}{%
\begin{tabular}{lrrrrrr|rrrrrr|rrrrrr}
\toprule
 & \multicolumn{6}{c}{Claude (haiku)} & \multicolumn{6}{c}{GPT (4o-mini)} & \multicolumn{6}{c}{Gemini (2.5-flash-lite)} \\ \cmidrule{1-19}
 & \multicolumn{2}{c}{Gemma 2 9B} & \multicolumn{2}{c}{Llama 3.1 8B} & \multicolumn{2}{c}{Mistral 7B} & \multicolumn{2}{c}{Gemma 2 9B} & \multicolumn{2}{c}{Llama 3.1 8B} & \multicolumn{2}{c}{Mistral 7B} & \multicolumn{2}{c}{Gemma 2 9B} & \multicolumn{2}{c}{Llama 3.1 8B} & \multicolumn{2}{c}{Mistral 7B} \\ \midrule
 \multicolumn{1}{l|}{Task ID} & \name & Rand. & \name & Rand. & \name & Rand. & \name & Rand. & \name & Rand. & \name & Rand. & \name & Rand. & \name & Rand. & \name & Rand. \\\midrule \multicolumn{1}{l|}{\odd}
& 165.5& 69.9 & 123.7& 69.9 & 79.8 & 69.9 & 103.4& 103.3& 242.1& 103.3& 241.5& 103.3& 42.1 & 37.8 & 78.0 & 37.8 & 72.2 & 37.8 \\\multicolumn{1}{l|}{\gh}
& 280.3& 156.2& 284.0& 156.2& 201.0& 156.2& 260.3& 140.4& 246.3& 140.4& 184.5& 140.4& 329.6& 189.9& 328.3& 189.9& 232.5& 189.9 \\\multicolumn{1}{l|}{\json}
& 159.9& 135.0& 172.1& 135.0& 164.9& 135.0& 21.3 & 14.9 & 26.4 & 14.9 & 7.9  & 14.9 & 23.6 & 15.9 & 28.2 & 15.9 & 9.8  & 15.9 \\\multicolumn{1}{l|}{\model}
& 137.5& 99.7 & 132.6& 99.7 & 149.6& 99.7 & 220.4& 140.7& 199.9& 140.7& 225.3& 140.7& 298.6& 202.6& 289.6& 202.6& 315.2& 202.6 \\\multicolumn{1}{l|}{\pos}
& 126.3& 66.4 & 108.6& 66.4 & 111.9& 66.4 & 145.0& 81.7 & 137.8& 81.7 & 137.6& 81.7 & 114.4& 63.4 & 99.7 & 63.4 & 96.2 & 63.4 \\\multicolumn{1}{l|}{\spell}
& 187.4& 58.5 & 152.2& 58.5 & 136.0& 58.5 & 153.2& 50.5 & 128.9& 50.5 & 110.8& 50.5 & 173.1& 60.3 & 137.0& 60.3 & 119.1& 60.3 \\\multicolumn{1}{l|}{\syn}
& 111.1& 60.5 & 81.5 & 60.5 & 112.0& 60.5 & 321.2& 179.0& 201.7& 179.0& 287.2& 179.0& 143.5& 85.1 & 114.4& 85.1 & 136.0& 85.1 \\\multicolumn{1}{l|}{\topic}
& 139.4& 79.3 & 122.6& 79.3 & 110.2& 79.3 & 163.4& 86.9 & 133.8& 86.9 & 128.9& 86.9 & 170.0& 94.9 & 149.3& 94.9 & 141.6& 94.9 \\
\bottomrule
\end{tabular}
} %
\end{table}

\begin{rqbox}[Answer to RQ4]
Adequacy scores that \name learns from open-weight models can be successfully transferred to closed-weight LLMs, enabling effective test prioritisation of failure-inducing inputs without any access to the internals of proprietary LLMs.
\end{rqbox}

\section{Discussion}
We discuss implications of our findings and outline directions for future work.
\label{sec:discussion}

\subsection{Distinguishing Different Failure Modes}

Our evaluation focuses on \name's ability to identify inputs that trigger failures, aligning with the core aim of robustness testing: prioritising failure-inducing cases. Since failures are relatively common in LLMs (as confirmed in our studied tasks), testers may also want to distinguish among \emph{types} of failures. For example, in syntactic bug detection, it is often more valuable to gather one instance from several bug categories (e.g., missing brackets, incorrect keywords) than to collect many instances of the same type.

Since \name partitions the input space with a GMM, its components may reveal distinct modes of failure. Targeting low-likelihood inputs from different components could increase the diversity of failures found. Beyond this, building models of failing inputs, potentially one per failure category, alongside the passing-input model offers another path to sharpen failure discovery and characterisation.

Another avenue is to distinguish adversarial inputs—\emph{deliberately} crafted to elicit unexpected behaviour—from non-adversarial yet failure-inducing ones. While our evaluation focuses on non-adversarial, task-valid cases, early evidence shows adversarial inputs produce sharply divergent representations with much lower likelihoods, suggesting \name could help separate the two.

\subsection{Beyond Binary Correctness and Simple Prompts}

Tasks in our study, like many existing LLM benchmarks, adopt binary correctness criteria. Yet real-world applications (e.g., counselling chatbot, AI language tutor) often demand richer evaluation, with graded or multi-dimensional measures, such as tone appropriateness, factual accuracy, or consistency. Recent evaluation frameworks~\cite{promptfoo, deepeval, giskard, langsmith} support such score-based assessments, typically assigning continuous scores (e.g., 0-1) using LLM-as-a-judge or human annotators.

We account for non-determinism through majority-pass criterion over repeated runs, and this idea of setting a \emph{passing} boundary for reference inputs can naturally extend to thresholding the continuous scores. Here, two design choices emerge: (i) aggregating scores across multiple dimensions to construct a unified reference set that satisfy overall quality requirements, or (ii) constructing separate reference sets for each evaluation criterion (e.g., \emph{tone-appropriate} vs., \emph{factually-accurate} passing sets). This perspective raises an interesting research question: whether transformer internal states encode signals that distinguish between a certain property-following and property-violating behaviours. If such signals exist, input adequacy could be redefined to align with finer-grained testing goals beyond binary correctness.

Modern LLM use also involves complex interaction patterns, such as tool use in autonomous agents, where correctness unfolds across multiple reasoning steps and each step has distinct failure modes. For instance, a tool-use agent may fail by (1) invoking a tool with invalid arguments, (2) misinterpreting tool outputs, or (3) reasoning incorrectly after the tool call. Applying \name step by step, estimating adequacy from internal representations before each action, could help predict failures early and guide targeted interventions. An open question is whether \name can scale to full agent trajectories spanning multiple conversation turns, and how the task-specific internal space shifts over the course of execution.

This paper studies the input adequacy for LLMs in the context of relatively simple tasks and prompts, as we believe such prompts are building blocks of more complicated LLM based systems. Future work can consider prompts designed for longer inferences and trajectories, including various agents designed for software engineering tasks~\cite{kang2024quantitative, zhang2024autocoderover, bouzenia2025repairagent}.

\subsection{Alternative Layer Depths}
Within the scope of our study, we selected a middle-to-late layer (2/3 depth) guided by the separability observed in the t-SNE visualisation (Section~\ref{sec:exploratory_token_layer_choice}). To further validate this choice, we additionally evaluate representations from earlier layers (1/3 and 1/2 depths) under the same labelling budget. Overall, we observe that performance degrades as we move away from the selected depth, with the correlation decreasing by 38.8\% at 1/3 depth and by 12.5\% at 1/2 depth on average. Looking more closely at individual models, we also find that the effective layer depth can vary: while Gemma tends to perform better at deeper layers, Llama and Mistral exhibit comparable performance even at half depth.

These results provide partial support for our initial assumption that middle-to-late layers yield more informative signals for difficulty estimation. At the same time, they suggest that further analysis of layer selection across different models and tasks may be beneficial, potentially requiring additional tuning or model-specific investigation.

\section{Threats to Validity}
\label{sec:threats}

Internal validity could be threatened because LLMs are not fully interpretable, meaning that \name might capture latent patterns that correlate with but are not identical to task-specific difficulty; using open-weight models may improve transparency and reproducibility but does not resolve this gap. External validity could also be at risk since \name learns task- and model-specific input distributions, which might limit generalisation, especially if other architectures such as reasoning LLMs (e.g., DeepSeek R1~\cite{guo2025deepseek}) produce different latent spaces; we attempt to reduce this risk by evaluating multiple models and tasks, though broader studies will be useful. Construct validity could likewise be threatened if our chosen metrics do not fully reflect adequacy, and while we use established measures (ROC-AUC, APFD) and a scrutinised, open-source GMM implementation~\cite{Pedregosa2011tk}, these proxies may not perfectly represent the underlying concept.

\section{Related Work}
\label{sec:related_work}

\noindent\textbf{Post-Generation Uncertainty:} LLM outputs have been widely studied to measure uncertainty and, consequently, detect hallucinations. Chen et al. propose EigenScore, which leverages the eigenvalue decomposition of activation covariance matrices to detect hallucinations ~\cite{chen2024inside}. The MIND approach of Su et al. goes further and toward online use, showing that unsupervised detectors over streaming internal states can flag hallucinations in real time during decoding~\cite{SuEtAl2024unsupervised}. Factoscope from He et al. aggregates multi-layer inner states with a Siamese network architecture to separate factual from non-factual outputs across models, reinforcing that hidden-state geometry correlates with veracity~\cite{HeEtAl2024LLM}. \name, on the other hand, uses only the internal hidden space of inputs.

\noindent\textbf{Input-side Internal Activation:} Recent work show that an LLM's input-side internal activations already contain strong signals about errors in the output. Azaria and Mitchell train a lightweight classifier on hidden states to distinguish true statements from the false, outperforming probability-only baselines and demonstrating that pre-generation representations encode truthfulness cues~\cite{azaria-mitchell-2023-internal}. Snyder et al. probe tokens and activations before the hallucination begins and train binary detectors that can forecast hallucinations in factual QA~\cite{SnyderMZ2024}. Extending this, Kossen et al.~\cite{kossen2024semantic} introduce Semantic Entropy Probes (SEPs) that approximate semantic entropy from a single forward pass (vs. multi-sample SE~\cite{farquhar2024detecting}), achieving strong detection and improving out-of-distribution generalisation. \name also uses the input-side activations, but measures task-specific test adequacy via reference set construction.

\noindent\textbf{Out-of-distribution Detection:} For deep neural networks, proposed hidden-state OOD detectors assess how atypical an input is relative to training data. Distance-based methods model class-conditional Gaussians and flag inputs with large Mahalanobis distance~\cite{lee2018simple}. Density-based methods fit mixture models over utterance embeddings, such as GMMs, for unknown intent detection~\cite{yan-etal-2020-unknown}. 
Reconstruction-based methods rely on hierarchical VAEs~\cite{havtorn2021hierarchical}. 
Unlike these approaches, which centre on detecting outliers or adversarial cases with respect to training distributions, \name addresses task-specific test adequacy by adaptively building a labelled reference set and modelling pre-generation hidden states in LLMs.

\section{Conclusion}
\label{sec:conclusion}

We propose \name, a technique that measures task-specific test adequacy for LLM inputs at the pre-generation stage. It iteratively constructs a set of reference inputs, whose internal latent space captures the distribution of inputs that are likely to pass. \name uses modelling and an adaptive sampling strategy to reduce the number of reference set inputs that need to be labelled by humans. The learnt distribution is subsequently used to compute task-specific Surprise Adequacy scores that can rank unseen test inputs.

Our empirical evaluation, across diverse tasks, show that \name accurately models the latent distributions of inputs that are likely to pass, and can achieve effective test prioritisation performance. Importantly, adequacy scores learnt on open-weight LLMs transfer effectively to proprietary models, allowing for cost-efficient testing without repeated and expensive LLM inferences.

Future work include evaluation against a wider range of both open and proprietary models, as well as investigating alternative sampling strategies that can further reduce labelling efforts. More broadly, we see pre-generation adequacy metrics as a promising foundation for scalable, task-aware testing of LLM-based systems, as well as hallucination and adversarial input detection.

\section*{Data Availability}
\label{sec:data_availability}

The data for the studied tasks, as well as the code for replication of the findings of this paper, are publicly available from GitHub\footnote{\url{https://github.com/coinse/clotho}} and Zenodo~\cite{clotho2026artifact}.

\section*{Acknowledgments}
\label{sec:ack}

Juyeon Yoon, Somin Kim, and Shin Yoo are supported by the National Research Foundation of Korea grant (RS-2026-25470171), the Engineering Research Center Program (RS-2021-NR060080), and the Institute of Information \& Communications Technology Planning \& Evaluation (IITP) grant (RS-2022-II220995), all funded by the Korean Government (MSIT). Robert Feldt is supported by Wallenberg AI, Autonomous Systems, and Software Program (WASP, BoundMiner project), the Swedish Science Foundation (No. 2020-05272, AQUAS) and the Chalmer's Foundation's Academic Excellence Program (CAEP-2026).

\bibliographystyle{ACM-Reference-Format}
\bibliography{ref}

\end{document}